\documentclass[11pt,a4paper]{article}
\usepackage{jheppub}
\usepackage{stmaryrd}

\newcommand{\g}{{\mathfrak g}}
\newcommand{\h}{{\mathfrak h}}
\newcommand{\m}{{\mathfrak K}}
\newcommand{\Y}{{\rm Y}}
\newcommand{\Kappa}{\text{\Large$\kappa$}}
\newcommand{\be}{\begin{equation}}
\newcommand{\ee}{\end{equation}}

\usepackage{tikz}\usetikzlibrary{matrix,arrows,shapes,decorations.markings}
\newcommand{\btp}{\begin{tikzpicture}[baseline=-5pt,scale=0.25,line width=0.7pt]}
\newcommand{\etp}{\end{tikzpicture}}

\newcommand{\twobv}{\begin{array}{c}\boxslash\vspace{-.19cm}
\end{array}\hspace{-0.44cm}\begin{array}{c}\boxslash\vspace{.25cm}\end{array}\!}

\title{Achiral boundaries and the twisted Yangian \\of the D5-brane}

\author[a]{Niall MacKay}
\author[a,b]{and Vidas Regelskis}

\affiliation[a]{Department of Mathematics, University of York,\\Heslington, York YO10 5DD, UK}
\affiliation[b]{Institute of Theoretical Physics and Astronomy of Vilnius University,\\Go\v{s}tauto 12, Vilnius 01108, Lithuania}

\emailAdd{niall.mackay@york.ac.uk}
\emailAdd{vr509@york.ac.uk}

\abstract{We consider integrable field theories with achiral boundary conditions and uncover
the underlying achiral twisted Yangian algebra. This construction arises from old work on the bosonic principal chiral model on a half-line, but finds a modern realization as the hidden symmetry in the planar limit of the scattering of worldsheet excitations of the AdS/CFT light-cone superstring off a $D5$-brane.}

\begin{document}

\maketitle


\section{Introduction}

Over the last ten years, the mature field of integrable models has found profound new applications
in the AdS/CFT correspondence \cite{overview}. The sub-field of the construction of
integrable boundaries has been similarly productive \cite{overviewDOOB}, but the
hidden symmetry it leads us to expect in AdS/CFT $D$-branes has not yet been fully revealed.

In particular, the most intensively studied integrable boundaries in AdS/CFT are
the $D3$ and $D7$-branes which, from the light-cone scattering theory point of view,
are of  chiral (chirality-preserving) type \cite{HF,CY,MR1,BP,AC}. It turns out \cite{AN,MR2,LP}
that these have a hidden symmetry consisting of a boundary twisted Yangian of the same
type as was found some years ago in the purely-bosonic principal chiral model (PCM;
the $G\times G$-invariant non-linear sigma model) on the half-line \cite{DMS03}, and
that this is, as expected for a boundary symmetry, a co-ideal subalgebra of the Yangian
of the AdS/CFT bulk $S$-matrix \cite{BeisertY,AT1,AT2}.

A second, achiral (chirality-reversing) type of integrable boundary conditions
is possible in the PCM, but was not explored in \cite{DMS03,MS03}. However, recent
study of reflection from the $D5$-brane \cite{CRY} by means of the coordinate
Bethe ansatz has shown that it corresponds precisely to reflection of this achiral type.
This raises the question of whether there is some form of hidden achiral boundary
Yangian symmetry, both in the PCM and for the $D5$-brane.

This paper therefore has two tasks. The first is to investigate the achiral boundary
conditions of the bosonic PCM, and to uncover the boundary remnant of the bulk
$\Y(\g \times \g)$ symmetry, demonstrating its coideal subalgebra property.
The second is to uncover the hidden Yangian symmetry of reflection from the $D5$-brane,
which turns out to be of exactly this type.

It is striking that, in all the ($D3, 5$ and $7$) cases, the coideal property governs
reflection, and  that the boundary symmetry is a generalized twisted Yangian, in which
the bulk and boundary Lie algebra symmetry form a symmetric pair. Further, we have now
realized examples of both classes of compact  symmetric space: $G/H$, where $H$ is
an involution-invariant subgroup, in the $D3$ and $D7$ cases, and $G\times G/G$ in the $D5$ case.

The paper is organized as follows. In section two, we consider the bosonic principal chiral model
defined on the half-line. We recall the achiral boundary conditions and uncover the remnant of
the bulk $\Y(\g \times \g)$ Yangian symmetry which they preserve, and which we call the
`achiral twisted Yangian', written $\Y(\g \times \g, \g)$ in the scheme of \cite{M02}.
In section three, we  briefly review the Yangian symmetry
of the AdS/CFT $S$-matrix constructed in \cite{BeisertY} and reflection from the
$D5$-brane, following closely \cite{CRY}. We then construct the achiral twisted
Yangian of the $D5$-brane for both `horizontal' and `vertical' orientations,
in folded and unfolded pictures of the reflection, and explore its building
blocks in detail.

\section{Achiral boundary conditions in the bosonic Principal Chiral Model}

Some ten years ago one of us considered \cite{MS03} the half-line $x\leq 0$ boundary
conditions which preserve the integrability of the $1+1$-dimensional bosonic principal chiral field $g(t,x)\in G$, for a compact, simple Lie group $G$, with Lagrangian density
\be\label{pcmlagr}
{\cal L} = \frac{1}{2} {\rm Tr}\left( \partial_+ g^{-1}
\partial_- g\right) \, .
\ee
The model has Lie algebra $\g$-valued conserved currents 
\be
j^L_\mu = \partial_\mu g g^{-1} \qquad {\rm and} \qquad j^R_\mu = - g^{-1}\partial_\mu g,
\ee 
which generate $\g_L$ and $\g_R$ and thereby the model's $G\times G$ symmetry.
On these currents at $x=0$, the boundary conditions  are either
\textit{chiral},
\be\label{chiral}
j_\pm^L=\alpha(j_\mp^L)\,, \quad j_\pm^R=\alpha(j_\mp^R)\quad
  \Rightarrow \quad j_0=\alpha(j_0)\,,\quad j_1=-\alpha(j_1) \quad {\rm (both} \,L\; {\rm and} \;R\,),
\ee
or \textit{achiral},\footnote{Formerly, in \cite{MS03}, `non-chiral'.
This author has since married a classicist, and now knows better than to mix Greek and Latin etymologies.}
\be\label{achiral}
j_\pm^L=\alpha(j_\mp^R) \quad \Rightarrow \quad j_0^L=\alpha(j_0^R)\,,\quad j_1^L=-\alpha(j_1^R),
\ee
where $\alpha$ is an involutive automorphism of $\g$.

For the chiral conditions the residual Lie symmetry is $H\times H\subset G\times G$,
where $H$ is the subgroup fixed by $\alpha$. In fact (\ref{chiral}) may be generalized
by independent conjugation of the currents, and the general boundary condition on the
fields $g$ is that
\be
g(t,0)\in k_L H k_R^{-1},
\ee
so that at $x=0$
\be
k_L^{-1} j_+^L k_L = \alpha( k_L^{-1} j_-^L k_L),\qquad k_R^{-1} j_+^R k_R = \alpha(k_R^{-1} j_-^L k_R).\label{chiral2}
\ee
The constant group elements $k_L$ and $k_R$ parametrize left- and right-cosets of
$H$ in $G$ and may be taken to lie in the Cartan immersion,
$G/H = \{ \alpha(k)k^{-1}\,|\, k\in G\}$, of $G/H$ in $G$.\footnote{Here $=$ means
`is locally diffeomorphic to'; there may be global ambiguities \cite{AP}.}

The Yangian symmetry is two ($L$ and $R$) copies of a generalization of
the twisted Yangian, $\Y(\g,\h)\subset \Y(\g)$ \cite{DMS03,M02}.
Writing $\g=\h\oplus\m$, one has
\begin{equation}
\left[\h,\h\right]\subset\h,\qquad\left[\h,\m\right]\subset\m,
\qquad\left[\m,\m\right]\subset\h,\label{symmetric_pair}
\end{equation}
respecting the eigenvalues $\alpha(\h)=+1$ and $\alpha(\m)=-1$, and
orthogonality with  respect to the Killing form, $\kappa(\h,\m)=0$.
This is crucial in guaranteeing that $\Y(\g,\h)$ is a co-ideal subalgebra: that the coproduct
of any Yangian charge $\mathbb{J}$ preserved by the boundary
lies in the tensor product of bulk and boundary Yangian
\begin{equation}
\Delta\mathbb{J}\in\Y(\g)\otimes\Y(\g,\h).\label{coideal}
\end{equation}
The co-ideal property ensures that multiparticle products of bulk and boundary states
represent $\Y(\g,\h)$, and is analogous to the requirement that $\Delta$ be a homomorphism
$\Y(\g)\rightarrow\Y(\g)\otimes\Y(\g)$ imposed by multiparticle bulk states.
If we let $i(,j,k,...)$ run over the $\h$-indices and $p,q(,r,...)$
over the $\m$-indices, we find that $\Y(\g,\h)$
is generated at level-0 by the usual $\h$ generators, $\mathbb{J}^{i}$, but at level-1,
in order that the co-ideal property hold, by a quadratic deformation of
the usual $\m$ generators $\hat{\mathbb{J}}^{p}$, which we write as
\begin{equation}
\tilde{\mathbb{J}}^{p} := \hat{\mathbb{J}}^{p}+\frac{1}{4}f_{\;\, qi}^{p}\,(\mathbb{J}^{q}\,\mathbb{J}^{i}+ \mathbb{J}^{i}\,\mathbb{J}^{q}).\label{twist}
\end{equation}
For different choices of $\h$ this encompasses twisted Yangians \cite{Ol92} and
reflection algebras \cite{Molev01}, and the `soliton-preserving' and
`-non-preserving' boundary conditions of \cite{Doikou}.

The second, achiral class of boundary condition was not fully investigated in \cite{MS03}.
Its Lie symmetry was under-identified there as the diagonal $H\subset G\times G$, but the full symmetry
is a diagonal $G\subset G\times G$, due to the conservation (where $\mathbb{J}:=\int_{-\infty}^0 j_0$)
\be
\frac{d}{dt} \left( \mathbb{J}_L + \alpha(\mathbb{J}_R) \right)
  = \int_{-\infty}^0 \partial_0 j_0^L + \partial_0 \alpha(j_0^R) = j_1^L(0) + \alpha(j_1^R(0)) =0.
\ee
Again conjugation is allowed, and
\be
\label{achiral2} g_L^{-1} j_+^L g_L = \alpha (g_R^{-1} j_-^R g_R)
\ee
at $x=0$ follows from
\be\label{Cartan}
g(t,0)\in g_L \{ \alpha(k)k^{-1}\,|\, k\in G\} g_R^{-1} = g_L  \,G/H \, g_R^{-1}.
\ee
Henceforth we set $g_L=g_R=1$ (the identity in $G$) for simplicity.

What is the remnant of the Yangian symmetry $\Y(\g \times \g)$? One might at first think
that it is simply $\Delta \Y(\g)$, but it is not. Rather it is again associated with
a symmetric space structure, this time $G\times G / G$, and is the co-ideal subalgebra
$\Y(\g\times \g,\g)$. This is expected from \cite{MY04}, which analysed boundary conditions
for symmetric-space sigma models.
We write the symmetric pair structure as $\g_L \oplus \g_R = \g_+ \oplus \g_-$, where $\g_+$
is the $\alpha$-twisted diagonal subalgebra and $\g_-$ its complement (which is {\em not} a
Lie algebra). Note especially that in this case different choices of $\alpha$ merely give
different $\alpha$-twisted embeddings of $\g$ in $\g \times \g$, rather than the different
proper subalgebras $\h$ of $\g$ we saw in the chiral case. Thus by a change of basis for $\g_R$ 
we can set $\alpha=\rm{id}$ (the identity map), and indeed we shall need such a change in 
the next section. For the moment we retain $\alpha$, but the reader may
like to bear in mind that $\alpha=\rm{id}$ captures the essence of the
construction.

The subalgebra $\g_+$ and its complement $\g_-$ are spanned by
$\mathbb{J}^a_\pm =\mathbb{J}^a_L \pm \alpha(\mathbb{J}^a_R)$, which have
 eigenvalues $\pm 1$ under the involution $\sigma(\alpha \times \alpha)$. The boundary
Yangian symmetry $Y(\g\times\g,\g)= Y(\g_L\times \g_R,\g_+)$, which we call the `achiral
twisted Yangian', then has Lie subalgebra $\g_+$, generated by the $\mathbb{J}_+^a$.
At level 1 its generators $\tilde{\mathbb{J}}_-^{a}$ are constructed from the level-1
$Y(\g_L\times \g_R)$ generators $\hat{\mathbb{J}}^a_L,\,\hat{\mathbb{J}}^a_R$ as
\begin{eqnarray}
\tilde{\mathbb{J}}_{-}^{a} & := & \hat{\mathbb{J}}_{-}^{a}+\frac{1}{8}f_{\;\, cb}^{a}\,(\mathbb{J}_{-}^{c}\,
\mathbb{J}_{+}^{b}+\mathbb{J}_{+}^{b}\,\mathbb{J}_{-}^{c})\label{twist_d}\nonumber\\
 & \;= & \hat{\mathbb{J}}_{-}^{a}+\frac{1}{2}f_{\;\, bc}^{a}\,
\mathbb{J}_L^{b}\,\alpha(\mathbb{J}_R^{c}),\label{twist2}
\end{eqnarray}
where again $\hat{\mathbb{J}}^a_\pm =\hat{\mathbb{J}}^a_L \pm \alpha(\hat{\mathbb{J}}^a_R)$.
Notice the factor of two in (\ref{twist_d}) relative to (\ref{twist}), due to the normalization of $\mathbb{J}^a_\pm$.

It is an easy calculation to check that these charges are classically conserved
by the achiral boundary condition (\ref{achiral}). Further, the co-product of the level-1 charges is
\begin{eqnarray}
\Delta\tilde{\mathbb{J}}_{-}^{a} & = & \Delta\hat{\mathbb{J}}_{-}^{a}+\frac{1}{8}f_{\; \, cb}^{a}(\Delta\mathbb{J}_{+}^{b}\,\Delta\mathbb{J}_{-}^{c}+\Delta\mathbb{J}_{-}^{c}\,\Delta\mathbb{J}_{+}^{b})\nonumber \\
& = & \hat{\mathbb{J}}_{-}^{a}\otimes1+1\otimes\hat{\mathbb{J}}_{-}^{a}+\frac{1}{8}f_{\; \, cb}^{a}(\mathbb{J}_{+}^{b}\,\mathbb{J}_{-}^{c}+\mathbb{J}_{-}^{c}\,\mathbb{J}_{+}^{b})
\otimes1+\frac{1}{8}f_{\;\, cb}^{a}\,1\otimes(\mathbb{J}_{+}^{b}\,\mathbb{J}_{-}^{c}+\mathbb{J}_{-}^{c}\,
\mathbb{J}_{+}^{b})\nonumber \\
&  & +\frac{1}{4}f_{\;\, bc}^{a}\left(\mathbb{J}_{-}^{b}\otimes\mathbb{J}_{+}^{c}+
\mathbb{J}_{+}^{b}\otimes\mathbb{J}_{-}^{c}\right)+\frac{1}{4}f_{\;\, cb}^{a}\left(\mathbb{J}_{+}^{b}\otimes\mathbb{J}_{-}^{c}-
\mathbb{J}_{-}^{b}\otimes\mathbb{J}_{+}^{c}\right)\nonumber \\
& = & \tilde{\mathbb{J}}_{-}^{a}\otimes1+1\otimes\tilde{\mathbb{J}}_{-}^{a}+\frac{1}{2}f_{\; \, bc}^{a}\,\mathbb{J}_{-}^{b}\otimes\mathbb{J}_{+}^{c} \label{co-ideal_d}\label{coideal2}\\
& \in & \Y\left(\g_L\times\g_R\right)\otimes\Y\left(\g_L\times\g_R,\g_+\right),\nonumber
\end{eqnarray}
and thus satisfies the crucial co-ideal property, analogous to (\ref{coideal}).

We commented earlier that the structure of $\Y(\g_L \times \g_R,\g_+)$ is subtly different
from a simple diagonal embedding $\Delta \Y(\g_+) \subset \Y(\g_L\times \g_R)$. This difference can be made precise, and is seen in the algebra isomorphism
\be
\Y(\g_L\times\g_R,\g_+) \cong \tilde\Delta\Y(\g)\label{YYre}.
\ee
Here, for any Yangian charge $\mathbb{J}$, we define
$\tilde\Delta\mathbb{J}:= (1\circ(-1)^l\alpha)\Delta\mathbb{J}$,
where $l$ is the level of $\mathbb{J}$, and $\Delta\mathbb{J}$ acts on
the tensor product of $L$ and $R$ components, written using a circle $\circ$.
Thus the difference lies entirely in the twisting by $1\circ(-1)^l\alpha$,
and at its simplest, with $\alpha=$id, solely in some relative signs.
At level 0 the isomorphism relates $\mathbb{J}^a_\pm$ on the left of (\ref{YYre}) to
$\mathbb{J}^a\circ 1 \pm 1 \circ \alpha(\mathbb{J}^a)$ on the right,
and at level 1 it relates $\tilde{\mathbb{J}}^a_-$ to
$\hat{\mathbb{J}}^a\circ 1 - 1 \circ \alpha(\hat{\mathbb{J}}^a )$.

At the co-algebra level this isomorphism takes the form
\be
\Delta\Y(\g_L\times\g_R,\g_+)\cong
\Sigma\cdot\Bigl(\Delta\circ\Delta'\bigl(\tilde\Delta\Y(\g)
\bigr)\Bigr).\label{YYrel}\ee
This equation is valued in the fourfold product of $\Y(\g)$ which acts on
the $L\circ R$ components of a $bulk\,\otimes\,boundary$ state; $\Delta$
and $\Delta':= \sigma\cdot\Delta$ are the usual and flipped coproducts. The role
of the operator $\Sigma$ is simply to re-arrange the order of factors,
$x_1\otimes x_2 \circ x_3 \otimes x_4 \mapsto x_1\circ x_4 \otimes x_2 \circ x_3 $.
The relations (\ref{YYre}, \ref{YYrel}) then capture the `folding' of a bulk
into a boundary scattering process \cite{CRY} which we shall meet shortly.

Let us demonstrate (\ref{YYrel}) explicitly. First,
\begin{eqnarray}
\Delta\circ\Delta'\bigl(\tilde\Delta\hat{\mathbb{J}}^a\bigr)
  & = & \Delta \circ \Delta'\left( \hat{\mathbb{J}}^a \circ 1 -  1\circ\alpha(\hat{\mathbb{J}}^a) +\frac{1}{2} f^a_{\;\,bc}\, \mathbb{J}^b \circ \alpha(\mathbb{J}^c)\right)\nonumber \\
& = & \left(\hat{\mathbb{J}}^a \otimes 1 + 1 \otimes \hat{\mathbb{J}}^a + \frac{1}{2} f^a_{\;\,bc}\, \mathbb{J}^b \otimes \mathbb{J}^c\right)\circ 1 \otimes 1 \nonumber \\
&& - \; 1 \otimes 1\circ \left(\alpha(\hat{\mathbb{J}}^a) \otimes 1 + 1 \otimes \alpha(\hat{\mathbb{J}}^a) - \frac{1}{2} f^a_{\;\,bc}\, \alpha(\mathbb{J}^b) \otimes \alpha(\mathbb{J}^c)\right)\nonumber \\
&& + \;\frac{1}{2} f^a_{\;\,bc} \left( \mathbb{J}^b \otimes 1  + 1 \otimes\mathbb{J}^b \right) \circ \left( {}^{}_{} \alpha(\mathbb{J}^c) \otimes 1  + 1 \otimes\alpha(\mathbb{J}^c) \right).
\end{eqnarray}
Then, acting with $\Sigma$,
\begin{eqnarray}
\Sigma\cdot\left( \Delta\circ\Delta'\bigl(\tilde\Delta\hat{\mathbb{J}}^a\bigr)\right) & = &
\left(\hat{\mathbb{J}}^a \circ 1 - 1\circ\alpha(\hat{\mathbb{J}}^a) +\frac{1}{2} f^a_{\;\,bc}\,\mathbb{J}^b\circ\alpha(\mathbb{J}^c)\right)\otimes 1 \circ 1 \nonumber \\&&
+ \; 1 \circ 1 \otimes \left(\hat{\mathbb{J}}^a \circ 1 - 1\circ\alpha(\hat{\mathbb{J}}^a) + \frac{1}{2} f^a_{\;\,bc}\,\mathbb{J}^b\circ\alpha(\mathbb{J}^c)\right)\nonumber \\&&
+ \; \frac{1}{2} f^a_{\;\,bc} \Bigl( \mathbb{J}^b\circ 1 - 1 \circ\alpha(\mathbb{J}^b)\Bigr) \otimes \Bigl(\mathbb{J}^c\circ 1 + 1 \circ\alpha(\mathbb{J}^c)\Bigr),
\end{eqnarray}
which, applying (\ref{twist2}), corresponds to (\ref{co-ideal_d}).

Now consider the implications for the scattering theory. Recall that the bulk multiplets
in the bosonic PCM form representations (hereafter `reps') $V\circ V$ of $\Y(\g_L\times \g_R)$,
where $V$ is a fundamental rep  of $\Y(\g)$.\footnote{This is typically a reducible rep of $\g$ with
the corresponding fundamental $\g$-rep as a component, although for $\g=\mathfrak{su}(n)$ they
are identical.} The bulk multiplet carries a rapidity $u$, corresponding to the application of
the shift automorphism $L_u\circ L_u$ to $Y(\g_L\times \g_R)$. The bulk scattering of
$U\circ U$ from $V\circ V$ is then constructed as a product of minimal factors $S_L\circ S_R$
(each factor acting on $U\otimes V$), multiplied by an overall scalar factor \cite{ORW}.\\

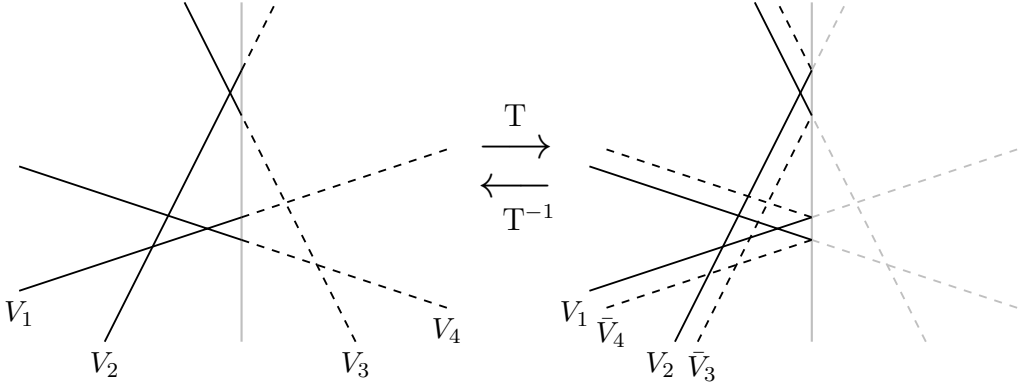
\begin{figure}[h] \begin{center} \begin{tikzpicture}[baseline=-5pt,scale=0.15,line width=0.7pt]
    \draw[lightgray,thick,-] (20,0) -- (20,30);
    \draw[-] (8,0) node[below] {$V_2$} -- (19,22) -- (15,30);
    \draw[-][dashed] (30,0) node[below] {$V_3$} -- (20,20);
    \draw[-] (20,20) -- (19,22) -- (20,24);
    \draw[-][dashed] (20,24) -- (23,30);
    \draw[-] (0.5,4.5) node[below] {$V_1$} -- (17,10) -- (0.5,15.5);
    \draw[-][dashed] (38,3) node[below] {$V_4$} -- (20,9);
    \draw[-] (20,9) -- (17,10) -- (20,11);
    \draw[-][dashed] (20,11) -- (38,17);

    \draw[lightgray,thick,-] (70,0) -- (70,30);
    \draw[-] (58,0) node[below] {$V_2\quad$} -- (69,22) -- (65,30);
    \draw[dashed] (60,0) node[below] {$\;\bar{V}_3$} -- (70,20);
    \draw[lightgray,dashed] (80,0) -- (70,20);
    \draw[-] (70,20) -- (69,22) -- (70,24);
    \draw[dashed] (70,24) -- (67,30);
    \draw[lightgray,dashed] (70,24) -- (73,30);
    \draw[-] (50.5,4.5) node[below] {$V_1\quad$} -- (67,10) -- (50.5,15.5);
    \draw[dashed] (52,3) node[below] {$\;\bar{V}_4$} -- (70,9);
    \draw[lightgray,dashed] (88,3) -- (70,9);
    \draw[-] (70,9) -- (67,10) -- (70,11);
    \draw[dashed] (70,11) -- (52,17);
    \draw[lightgray,dashed] (70,11) -- (88,17);

    \draw[black] (44,20) node {\large{T}} (44,17) node {\LARGE{$\longrightarrow$}}
    (45.2,11) node {\large{T$^{-1}$}} (44,13.5) node {\LARGE{$\longleftarrow$}};
\etp \\
\end{center}
\hspace{1.3in}\parbox[c]{3.4in}{\caption{The action of the conjugation operator T
on  4-particle scattering, with solid $L$ and dashed $R$ lines.}
\label{fig_Tfold}}
\end{figure}

The state $V\circ V$ scatters off the boundary into $\bar{V}\circ\bar{V}$, with $u\mapsto -u$. Thus, on the states, in the
isomorphism (\ref{YYre}) we write the action of $\Sigma$ as conjugation by an operator T
whose effect is to re-order and conjugate multiplets,
\be
\rm T : V_1\otimes V_2 \circ V_3 \otimes V_4 \rightarrow V_1 \circ \bar{V}_4 \otimes V_2 \circ \bar{V}_3.\label{T}
\ee
The meaning of the map T (T$^{-1}$) is revealed, as the folding (unfolding) of a bulk
to a boundary scattering process, in figure \ref{fig_Tfold}. Similar unfolding processes
relate boundary unitarity and crossing-unitarity \cite{GZ94} to bulk unitarity and
crossing relations. The reversing of rapidity by the boundary can be seen in how the
shift automorphism acts on (\ref{YYre}), where, crucially,
$L_u\circ L_u (1\circ(-1)^l\alpha)= (1\circ(-1)^l\alpha)L_u\circ L_{-u}$ on $\Delta\Y(\g)$.

\begin{figure}
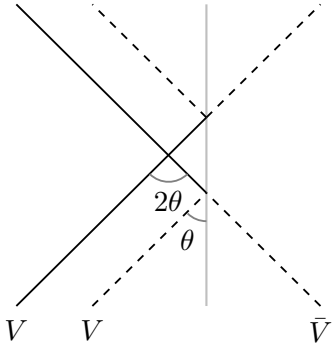

\begin{centering}
\btp
   \draw[lightgray,thick,-] (10,0) -- (10,16);
   \draw[-] (0,0) -- (10,10);
   \draw[-] (10,6)-- (0,16);
   \draw[-,dashed] (4,0) -- (10,6);
   \draw[-,dashed] (10,10) -- (4,16);
   \draw[-,dashed] (16,0)-- (10,6);
   \draw[-,dashed] (10,10) -- (16,16);
   \draw[gray] (9,7) arc (-45:-135:1.4);
   \draw[gray] (10,4.5) arc (-90:-135:1.4);
   \draw (8,5.6) node {$2\theta$} (9,3.5) node {$\theta$}
	  (0,-1.3) node {$V$} (4,-1.3) node {$V$} (16,-1.3) node {$\bar{V}$};
\etp
\par\end{centering}
\hspace{0.5in}%
\parbox[c]{5in}{%
\caption{Achiral reflection process.}
\label{fig_KS}}
\end{figure}

In the simplest case, where the boundary is in a singlet state, the boundary scattering
matrix $K_{V\circ V}(u)$ is conjugate to $(1\circ \alpha)  S_{V\bar{V}}(2u)$ (see
figure \ref{fig_KS}). Its direct construction via conservation of the
$\Y(\g_L\times \g_R,\g_+)$ charges is isomorphic, via (\ref{YYre}), to that of the bulk
S-matrix (using, for example, the Tensor Product Graph method \cite{M91}), in which the
doubling of $u$ is traced back to the extra factor of two in (\ref{twist_d}).
We therefore expect a spectrum of boundary bound states in non-trivial multiplets whose
mass ratios are those of the bulk states, inherited through the pole structure of
$S_{V\bar{V}}(2u)$.

\begin{figure}[h] \begin{center} \begin{tikzpicture}[baseline=-5pt,scale=0.2,line width=0.7pt]
    \draw[gray,thick,-] (11,-2) node [below] {$V_B\;$} -- (11,22);
    \draw[black] (5,0) node [below] {$V_2$} -- (17,20);
    \draw[black] (8,20) -- (20,0) node [below] {$V_3$};
    \draw[black] (0,2) node [below] {$V_1$} -- (25,12);
    \draw[black] (0,12) -- (25,2) node [below] {$V_4$};

    \draw[gray,thick,-] (46.5,-2) node [below] {$V_B\;$} -- (46.5,22);
    \draw[black] (37.5,0) node [below] {$V_2$} -- (49.5,20);
    \draw[black] (40.5,20) -- (52.5,0) node [below] {$V_3$};
    \draw[black] (32.5,2) node [below] {$V_1$} -- (57.5,12);
    \draw[black] (32.5,12) -- (57.5,2) node [below] {$V_4$};

    \draw (28.5,10) node {\LARGE{=}};
\etp \end{center}
\hspace{0.5in}\parbox{5.1in}{\caption{The unfolded bYBE as a 5-particle bulk process.
The vertical line corresponding to the boundary may be shifted left or right by
employing the bulk YBE.}\label{fig3}}
\end{figure}
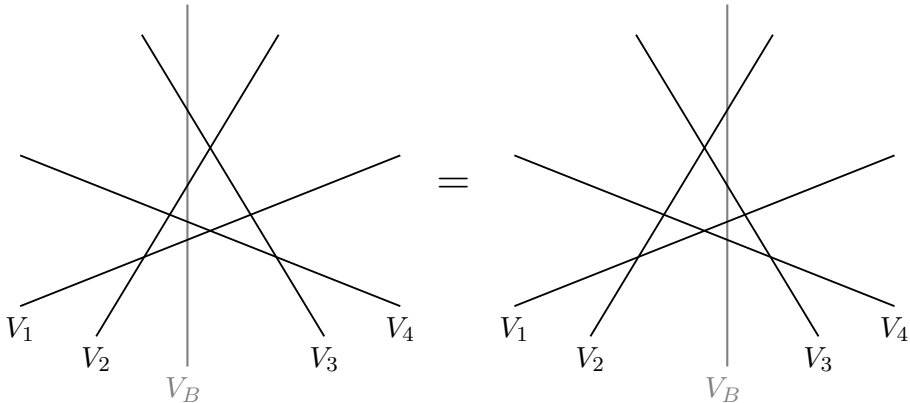

The folding construction of the boundary scattering process straightforwardly
accommodates such non-trivial boundary multiplets, and this will play an
important role in understanding reflection from the $D5$-brane in the next
section. The boundary scattering matrix, the relevant solution of the
boundary Yang-Baxter equation (bYBE) \cite{Cher,Sklyanin},  then becomes
a product of three non-trivial factors, analogous to the boundary fusion
procedure \cite{MN1,MN2}. These are a bulk $S$-matrix and two
(what we shall call) `achiral reflection matrices', which are trivial in
the case of the singlet boundary and which participate in the reflection
process as on the right of figure \ref{fig_Tfold}. This threefold process
inherits, via (\ref{YYre}), a Yang-Baxter property: the order of the
factorization does not matter, and our apparent placing of the bulk $S$-matrix
to the left of the boundary in figures \ref{fig_Tfold} and \ref{fig_KS}
is merely an artefact (figure \ref{fig3}).

\section{Yangian symmetry of the D5-brane}

In this section we briefly review the Yangian symmetry of the $S$-matrix presented in \cite{BeisertY}
and set up the scattering problem for reflection from the $D5$-brane, following closely \cite{CRY}.
We then combine this with the ideas of the preceding section to build
the Yangian symmetry of the $D5$-brane.

\subsection{Yangian symmetry of the bulk S-matrix}

The symmetry algebra of the excitations of the light-cone superstring
theory on $AdS_{5}\times S^{5}$ (and thereby of the $S$-matrix),
and also of the single trace operators in the $\mathcal{N}=4$ supersymmetric
gauge theory that are analogous to (and known as) spin chains, is generated
by the centrally-extended algebra
$\mathfrak{psu}(2|2)\times\widetilde{\mathfrak{psu}}(2|2)\ltimes\mathbb{R}^{3}$.
The magnons of AdS/CFT superstring transform in the reps
of this algebra and we shall call them the `left' and `right' reps
of the bulk magnon, and write the algebra as
$\mathfrak{psu}(2|2)_L\times\mathfrak{psu}(2|2)_R\ltimes\mathbb{R}^{3}$.

We will briefly review the $\mathfrak{psu}(2|2)\ltimes\mathbb{R}^{3}$
and its Yangian extension, since they are essential to the construction of
the Yangian of the $D5$-brane. The algebra has two sets of bosonic rotation generators
$\mathbb{R}_{a}^{\enskip b}$, $\mathbb{L}_{\alpha}^{\enskip\beta}$,
two sets of fermionic supersymmetry generators $\mathbb{Q}_{\alpha}^{\enskip a},$
$\mathbb{G}_{a}^{\enskip\alpha}$ and three central charges $\mathbb{H}$,
$\mathbb{C}$ and $\mathbb{C}^{\dagger}$. The nontrivial commutation
relations are \cite{BeisertS}
\begin{align}
 & \left[\mathbb{L}_{\alpha}^{\enskip\beta},\mathbb{J}_{\gamma}\right]=
 \delta_{\gamma}^{\beta}\,\mathbb{J}_{\alpha}-\frac{1}{2}\delta_{\alpha}^{\beta}\,\mathbb{J}_{\gamma},
 &  & \left[\mathbb{L}_{\alpha}^{\enskip\beta},\mathbb{J}^{\gamma}\right]=-
 \delta_{\alpha}^{\gamma}\,\mathbb{J}^{\beta}+\frac{1}{2}\delta_{\alpha}^{\beta}\,\mathbb{J}^{\gamma},
 \nonumber \\
 & \left[\mathbb{R}_{a}^{\enskip b},\mathbb{J}_{c}\right]=\delta_{c}^{b}\,\mathbb{J}_{a}-\frac{1}{2}\delta_{a}^{b}\,\mathbb{J}_{c}, &  & \left[\mathbb{R}_{a}^{\enskip b},\mathbb{J}^{c}\right]=-\delta_{a}^{c}\,\mathbb{J}^{b}+\frac{1}{2}\delta_{a}^{b}\,\mathbb{J}^{c},\nonumber \\
 & \left\{ \mathbb{Q}_{\alpha}^{\enskip a},\mathbb{Q}_{\beta}^{\enskip b}\right\} =\epsilon^{ab}\epsilon_{\alpha\beta}\,\mathbb{C}, &  & \left\{ \mathbb{G}_{a}^{\enskip\alpha},\mathbb{G}_{b}^{\enskip\beta}\right\} =\epsilon^{\alpha\beta}\epsilon_{ab}\,\mathbb{C}^{\dagger},\nonumber \\
 & \left\{ \mathbb{Q}_{\alpha}^{\enskip a},\mathbb{G}_{b}^{\enskip\beta}\right\} =\delta_{b}^{a}\,\mathbb{L}_{\beta}^{\enskip\alpha}+\delta_{\beta}^{\alpha}\,\mathbb{R}_{b}^{\enskip a}+\frac{1}{2}\delta_{b}^{a}\delta_{\beta}^{\alpha}\,\mathbb{H},\label{psu(2|2)_algebra}
\end{align}
where $a,\; b,...=1,\;2$ and $\alpha,\;\beta,...=3,\;4$.

The Yangian symmetry fixes the bound state $S$-matrices uniquely up to
an overall dressing phase \cite{dL,ALT} by requiring their invariance under the co-products
of the Yangian generators \cite{BeisertY}

\begin{align*}
\Delta\hat{\mathbb{R}}_{a}^{\enskip b} & =\hat{\mathbb{R}}_{a}^{\enskip b}\otimes1+1\otimes\hat{\mathbb{R}}_{a}^{\enskip b}+\frac{1}{2}\mathbb{R}_{a}^{\enskip c}\otimes\mathbb{R}_{c}^{\enskip b}-\frac{1}{2}\mathbb{R}_{c}^{\enskip b}\otimes\mathbb{R}_{a}^{\enskip c}-\frac{1}{2}\mathbb{G}_{a}^{\enskip\gamma}\otimes\mathbb{Q}_{\gamma}^{\enskip b}-\frac{1}{2}\mathbb{Q}_{\gamma}^{\enskip b}\otimes\mathbb{G}_{a}^{\enskip\gamma}\nonumber \\
 & \qquad+\frac{1}{4}\delta_{a}^{b}\mathbb{G}_{c}^{\enskip\gamma}\otimes\mathbb{Q}_{\gamma}^{\enskip c}+\frac{1}{4}\delta_{a}^{b}\mathbb{Q}_{\gamma}^{\enskip c}\otimes\mathbb{G}_{c}^{\enskip\gamma},\nonumber \\
\Delta\hat{\mathbb{L}}_{\alpha}^{\enskip\beta} & =\hat{\mathbb{L}}_{\alpha}^{\enskip\beta}\otimes1+1\otimes\hat{\mathbb{L}}
_{\alpha}^{\enskip\beta}-\frac{1}{2}\mathbb{L}_{\alpha}^{\enskip\gamma}\otimes
\mathbb{L}_{\gamma}^{\enskip\beta}+\frac{1}{2}\mathbb{L}_{\gamma}^{\enskip\beta}
\otimes\mathbb{L}_{\alpha}^{\enskip\gamma}+\frac{1}{2}\mathbb{G}_{c}^{\enskip\beta}
\otimes\mathbb{Q}_{\alpha}^{\enskip c}+\frac{1}{2}\mathbb{Q}_{\alpha}^{\enskip c}\otimes\mathbb{G}_{c}^{\enskip\beta}\nonumber \\
 & \qquad-\frac{1}{4}\delta_{\alpha}^{\beta}\mathbb{G}_{c}^{\enskip\gamma}\otimes\mathbb{Q}_{\gamma}^{\enskip c}-\frac{1}{4}\delta_{\alpha}^{\beta}\mathbb{Q}_{\gamma}^{\enskip c}\otimes\mathbb{G}_{c}^{\enskip\gamma},\nonumber \\
\Delta\hat{\mathbb{Q}}_{\alpha}^{\enskip a} & =\hat{\mathbb{Q}}_{\alpha}^{\enskip a}\otimes1+1\otimes\hat{\mathbb{Q}}_{\alpha}^{\enskip a}+\frac{1}{2}\mathbb{Q}_{\alpha}^{\enskip c}\otimes\mathbb{R}_{c}^{\enskip b}-\frac{1}{2}\mathbb{R}_{c}^{\enskip a}\otimes\mathbb{Q}_{\alpha}^{\enskip c}+\frac{1}{2}\mathbb{Q}_{\gamma}^{\enskip a}\otimes\mathbb{L}_{\alpha}^{\enskip\gamma}-\frac{1}{2}\mathbb{L}_{\alpha}^{\enskip\gamma}
\otimes\mathbb{Q}_{\gamma}^{\enskip a}\nonumber \\
 & \qquad+\frac{1}{4}\mathbb{Q}_{\alpha}^{\enskip a}\otimes\mathbb{H}-\frac{1}{4}\mathbb{H}\otimes\mathbb{Q}_{\alpha}^{\enskip a}+\frac{1}{2}\varepsilon_{\alpha\gamma}\varepsilon^{ad}\mathbb{C}\otimes\mathbb{G}_{d}^{\enskip\gamma}
 -\frac{1}{2}\varepsilon_{\alpha\gamma}\varepsilon^{ad}\mathbb{G}_{d}^{\enskip\gamma}\otimes\mathbb{C},\nonumber \\
\Delta\hat{\mathbb{G}}_{a}^{\enskip\alpha} & =\hat{\mathbb{G}}_{a}^{\enskip\alpha}\otimes1+1\otimes\hat{\mathbb{G}}_{a}^{\enskip\alpha}
-\frac{1}{2}\mathbb{G}_{c}^{\enskip\alpha}\otimes\mathbb{R}_{a}^{\enskip c}+\frac{1}{2}\mathbb{R}_{a}^{\enskip c}\otimes\mathbb{G}_{c}^{\enskip\alpha}-\frac{1}{2}\mathbb{G}_{a}^{\enskip\gamma}
\otimes\mathbb{L}_{\gamma}^{\enskip\alpha}+\frac{1}{2}\mathbb{L}_{\gamma}^{\enskip\alpha}
\otimes\mathbb{G}_{a}^{\enskip\gamma}\nonumber \\
 & \qquad-\frac{1}{4}\mathbb{G}_{a}^{\enskip\alpha}\otimes\mathbb{H}+\frac{1}{4}\mathbb{H}
 \otimes\mathbb{G}_{a}^{\enskip\alpha}-\frac{1}{2}\varepsilon_{ac}\varepsilon^{\alpha\gamma}
 \mathbb{C}^{\dagger}\otimes\mathbb{Q}_{\gamma}^{\enskip c}+\frac{1}{2}\varepsilon_{ac}\varepsilon^{\alpha\gamma}\mathbb{Q}_{\gamma}^{\enskip c}\otimes\mathbb{C}^{\dagger},\nonumber
\end{align*}
\begin{align}
\Delta\mathbb{C} & =\hat{\mathbb{C}}\otimes1+1\otimes\hat{\mathbb{C}}-\frac{1}{2}\mathbb{H}\otimes\mathbb{C}
+\frac{1}{2}\mathbb{C}\otimes\mathbb{H},\nonumber \\
\Delta\mathbb{C}^{\dagger} & =\hat{\mathbb{C}}^{\dagger}\otimes1+1\otimes\mathbb{\hat{C}}^{\dagger}+\frac{1}{2}\mathbb{H}
\otimes\mathbb{C}^{\dagger}-\frac{1}{2}\mathbb{C}^{\dagger}\otimes\mathbb{H},\nonumber \\
\Delta\mathbb{H} & =\hat{\mathbb{H}}\otimes1+1\otimes\hat{\mathbb{H}}+\mathbb{C}\otimes\mathbb{C}^{\dagger}
-\mathbb{C}^{\dagger}\otimes\mathbb{C}.\label{Y(g)}
\end{align}

The non-trivial braiding factors are not explicitly shown in the
co-products above---rather they are all hidden in the parameters of
the representation. As was shown in \cite{BeisertY}, the action of the
Yangian charges on the magnon states takes the form
\begin{equation}
\hat{\mathbb{J}}^{A}\left|u\right\rangle =\frac{i g}{2}u\,\mathbb{J}^{A}\left|u\right\rangle ,\label{uJ_ansatz}
\end{equation}
where $u=\frac{1}{2}(x^{+}+\frac{1}{x^{+}}-x^{-}-\frac{1}{x^{-}})$
is the rapidity of the state.


\subsection{The D5-brane: general considerations}

The $D5$-brane considered in \cite{CY,CRY} wraps an $AdS_{4}\subset AdS_{5}$
and a maximal $S^{2}\subset S^{5}$. Such a configuration defines a
$2+1$ dimensional defect hypersurface of the $3+1$
dimensional conformal boundary of $AdS_{5}$. The fundamental matter living on
this hypersurface is a 3d hypermultiplet \cite{DFO}.
The presence of the D5-brane breaks the $\mathfrak{so}(6)$ symmetry of $S^5$
down to $\mathfrak{so}(3)_{H}\times \mathfrak{so}(3)_{V}$.
As is the convention, we fix the bulk vacuum state
to be $Z=X^5+iX^6$ and then consider two different orientations of the maximal $S^{2}$ inside $S^5$:
\begin{itemize}
\item the maximal $S^{2}$ specified by $X^{1}=X^{2}=X^{3}=0$, -- this orientation in \cite{CRY} was
termed `horizontal' $D5$-brane\footnote{This configuration corresponds to horizontal
vacuum orientation in the scattering theory.} and, from the scattering theory point of view,
corresponds to a singlet boundary;
\item the maximal $S^{2}$ specified by $X^{4}=X^{5}=X^{6}=0$, -- this orientation is perpendicular to
the previous and is termed the `vertical' $D5$-brane; now the boundary carries a field multiplet
transforming in the vector representation of the boundary algebra.
\end{itemize}

\paragraph{The Lie algebra.} The $D5$-brane preserves a diagonal subalgebra
$\mathfrak{psu}(2|2)_{+}\ltimes\mathbb{R}^{3}$ of the bulk algebra
$\mathfrak{psu}(2|2)_L\times\mathfrak{psu}(2|2)_R\ltimes\mathbb{R}^{3}$
generated by 
\begin{align}
\mathbb{L}_{\check{\alpha}}^{\enskip\check{\beta}} & =\mathbb{L}_{\alpha}^{\enskip\beta}+\mathbb{L}_{\bar{\dot{\alpha}}}^{\enskip\bar{\dot{\beta}}}, &  & \mathbb{Q}_{\check{\alpha}}^{\enskip\check{a}}=\mathbb{Q}_{\alpha}^{\enskip a}+\kappa\,\mathbb{Q}_{\bar{\dot{\alpha}}}^{\enskip\dot{a}},\nonumber \\
\mathbb{R}_{\check{a}}^{\enskip\check{b}} & =\mathbb{R}_{a}^{\enskip b}+\mathbb{R}_{\dot{a}}^{\enskip\dot{b}}, &  & \mathbb{G}_{\check{a}}^{\enskip\check{\alpha}}=
\mathbb{G}_{a}^{\enskip\alpha}+\kappa^{-1}\mathbb{G}_{\dot{a}}^{\enskip\bar{\dot{\alpha}}},
\label{diag_symm}
\end{align}
where $\kappa=-1$ for the horizontal case and $\kappa=-i$ for the vertical one.\footnote{The
supersymmetries preserved by the $D5$-brane were worked out in Appendix B
of \cite{CY} and lead to two scattering theories with $\kappa^2=\pm1$, representing the horizontal and vertical cases.}
The notation for the dotted and checked indices is the same
as for undotted ones, $\dot{a},\,\check{a},\,\dot{b},\,\check{b}=1,\,2$
and $\dot{\alpha},\,\check{\alpha},\,\dot{\beta},\,\check{\beta}=3,\,4$.
The generators with the undotted indices
generate $\mathfrak{psu}(2|2)_L$ and the generators with the dotted
indices generate $\mathfrak{psu}(2|2)_R$. Rather than make the involution $\alpha$ explicit, it
is easier to absorb it into the combination of the scale $\kappa$ and a change of basis,
denoted by a bar, which acts on the dotted greek indices acts as $\bar{\dot{3}}=\dot{4}$
and $\bar{\dot{4}}=\dot{3}$.  We also
define the complementary charges
\begin{align}
\overline{\mathbb{L}}_{\check{\alpha}}^{\enskip\check{\beta}} & =\mathbb{L}_{\alpha}^{\enskip\beta}-\mathbb{L}_{\bar{\dot{\alpha}}}^{\enskip\bar{\dot{\beta}}},
  &  & \overline{\mathbb{Q}}_{\check{\alpha}}^{\enskip\check{a}}=\mathbb{Q}_{\alpha}^{\enskip a}-\kappa\,\mathbb{Q}_{\bar{\dot{\alpha}}}^{\enskip\dot{a}},\nonumber \\
\overline{\mathbb{R}}_{\check{a}}^{\enskip\check{b}} & =\mathbb{R}_{a}^{\enskip b}-\mathbb{R}_{\dot{a}}^{\enskip\dot{b}},
  &  & \overline{\mathbb{G}}_{\check{a}}^{\enskip\check{\alpha}}=\mathbb{G}_{a}^{\enskip\alpha}-\kappa^{-1}\mathbb{G}_{\dot{a}}^{\enskip\bar{\dot{\alpha}}},\label{comp_diag_symm}
\end{align}
which do not in themselves form a Lie algebra (and are not preserved by the boundary), but together with 
(\ref{diag_symm}) and the central charges $\mathbb{C}$, 
$\mathbb{C}^{\dagger}$ and $\mathbb{H}$ generate bulk algebra 
$\mathfrak{psu}(2|2)_L\times\mathfrak{psu}(2|2)_R\ltimes\mathbb{R}^{3}$.

\paragraph{The bulk representation.} The bulk magnon transforms
in the $\boxslash_{(a,b,c,d)}$ of the left and the $\widetilde{\boxslash}_{(a,b,c,d)}$
of the right rep of the bulk symmetry algebra and they both carry
the same set $(a,b,c,d)$ of representation labels. Our goal
is to build the canonical representation of bulk states with respect
to the boundary algebra \eqref{diag_symm}. It is easy to see that
the left rep transforms canonically with respect to the boundary
algebra. However, the right rep does not, and one thus has to
choose a different basis for it in order to obtain the algebra
action in the canonical form. It was shown in \cite{CRY}
that by choosing the basis
\begin{equation}
(\tilde{\phi}_{\check{1}},\tilde{\phi}_{\check{2}}|\tilde{\psi}_{\check{3}},\tilde{\psi}_{\check{4}})
  :=(\tilde{\phi}_{\dot{1}},\tilde{\phi}_{\dot{2}}|\lambda\tilde{\psi}_{\dot{4}},\lambda\tilde{\psi}_{\dot{3}})\label{new_basis}
\end{equation}
to be the new basis of $\widetilde{\boxslash}$, with some arbitrary
constant $\lambda$ representing the rescaling of the new base with
the respect to the old, one acquires the canonical action of the boundary
algebra (\ref{diag_symm}) on the right rep:
\begin{align}
\mathbb{Q}_{\check{\alpha}}^{\enskip\check{a}}\left|\tilde{\phi}_{\check{b}}\right\rangle  & =\tilde{a}\,\delta_{\check{b}}^{\check{a}}\left|\tilde{\psi}_{\check{\alpha}}\right\rangle , &  & \mathbb{G}_{\check{a}}^{\enskip\check{\alpha}}\left|\tilde{\phi}_{\check{b}}\right\rangle =\tilde{c}\,\varepsilon^{\check{\alpha}\check{\beta}}\,\varepsilon_{\check{a}\check{b}}\left|\tilde{\psi}_{\check{\beta}}\right\rangle ,\nonumber \\
\mathbb{Q}_{\check{\alpha}}^{\enskip\check{a}}\left|\tilde{\psi}_{\check{\beta}}\right\rangle  & =\tilde{b}\,\varepsilon_{\check{\alpha}\check{\beta}}\,\varepsilon^{\check{a}\check{b}}\left|\tilde{\phi}_{\check{b}}\right\rangle , &  & \mathbb{G}_{\check{a}}^{\enskip\check{\alpha}}\left|\lambda\tilde{\psi}_{\check{\beta}}\right\rangle =\tilde{d}\,\delta_{\check{\beta}}^{\check{\alpha}}\left|\tilde{\phi}_{\check{a}}\right\rangle ,\label{checked_action}
\end{align}
where $(\tilde{a},\tilde{b},\tilde{c},\tilde{d})$ are the representation
parameters in the new basis. They are related to the old basis by
\begin{equation}
\tilde{a}=\frac{\kappa}{\lambda}a,\quad\tilde{b} =-\kappa\lambda\, b,\quad\tilde{c}=-\frac{1}{\kappa\lambda}c,\quad\tilde{d}=\frac{\lambda}{\kappa}d,\label{abcd_rel}
\end{equation}
where the minus sign comes from the relation
$\varepsilon^{\bar{\dot{\alpha}}\bar{\dot{\beta}}}
 =\overline{\varepsilon^{\dot{\alpha}\dot{\beta}}}
 =\overline{\varepsilon^{\check{\alpha}\check{\beta}}}
 =-\varepsilon^{\check{\alpha}\check{\beta}}$
and similarly for $\varepsilon_{\bar{\dot{\alpha}}\bar{\dot{\beta}}}$.
By choosing $\tilde{a}=a$ one fixes the rescaling constant to be
$\lambda=\kappa$ and arrives at the relation between new and old
representation labels
\begin{equation}
(\tilde{a},\tilde{b},\tilde{c},\tilde{d})=(a,-\kappa^{2}b,-\kappa^{-2}c,d).\label{new_labels}
\end{equation}
Thus the canonical representation of the bulk magnon with respect
to the boundary algebra and the corresponding labels are
\begin{equation}
\boxslash_{(a,b,c,d)} \circ \; \widetilde{\boxslash}_{(a,-\kappa^{2}b,
-\kappa^{-2}c,d)}.\label{bulk_reps}
\end{equation}
As in the previous section, we use $\circ$ to denote the tensor product of $L$
and $R$ reps of bulk magnon, reserving the usual $\otimes$ for the
tensor product of the bulk and boundary reps. The action of the
complementary charges (\ref{comp_diag_symm}) on the right rep is
almost of canonical form, except for an extra minus sign
\begin{align}
\overline{\mathbb{Q}}_{\check{\alpha}}^{\enskip\check{a}}\left|\tilde{\phi}_{\check{b}}\right\rangle  & =-\tilde{a}\,\delta_{\check{b}}^{\check{a}}\left|\tilde{\psi}_{\check{\alpha}}\right\rangle , &  & \overline{\mathbb{G}}_{\check{a}}^{\enskip\check{\alpha}}\left|\tilde{\phi}_{\check{b}}\right\rangle =-\tilde{c}\,\varepsilon^{\check{\alpha}\check{\beta}}\,\varepsilon_{\check{a}\check{b}}\left|\tilde{\psi}_{\check{\beta}}\right\rangle ,\nonumber \\
\overline{\mathbb{Q}}_{\check{\alpha}}^{\enskip\check{a}}\left|\tilde{\psi}_{\check{\beta}}\right\rangle  & =-\tilde{b}\,\varepsilon_{\check{\alpha}\check{\beta}}\,\varepsilon^{\check{a}\check{b}}\left|\tilde{\phi}_{\check{b}}\right\rangle , &  & \overline{\mathbb{G}}_{\check{a}}^{\enskip\check{\alpha}}\left|\lambda\tilde{\psi}_{\check{\beta}}\right\rangle =-\tilde{d}\,\delta_{\check{\beta}}^{\check{\alpha}}\left|\tilde{\phi}_{\check{a}}\right\rangle .\label{hatted_action}
\end{align}
The total eigenvalues of the central charges on the bulk reps (\ref{bulk_reps})
in the new basis become
\begin{align}
\mathbb{C} & =\mathbb{C}\circ1+1\circ\mathbb{C}=ab+\tilde{a}\tilde{b}=(1-\kappa^{2})ab,\nonumber \\
\mathbb{C}^{\dagger} & =\mathbb{C}^{\dagger}\circ1+1\circ\mathbb{C}^{\dagger}=cd+\tilde{c}\tilde{d}=(1-\kappa^{-2})ab,\nonumber \\
\mathbb{H} & =\mathbb{H}\circ1+1\circ\mathbb{H}=ad+bc+\tilde{a}\tilde{d}+\tilde{b}\tilde{c}=2(ad+bc),\label{cc}
\end{align}
hence the bulk magnon lives in the following tensor product
of fundamental representations:
\begin{equation}
\left\langle 0,0;H,C,C^{\dagger}\right\rangle \circ\left\langle 0,0;H,-\kappa^{2}C,-\kappa^{-2}C^{\dagger}\right\rangle
 =\left\{ 0,0,2H,(1-\kappa^{2})C,(1-\kappa^{-2})C^{\dagger}\right\},
\end{equation}
which depends on the value of $\kappa$. Let us explain this result in more detail.

The symmetry algebra in the bulk is $\g_L \oplus \g_R \oplus \mathfrak{m}$,
where $\mathfrak{m}$ is the central extension, which is invariant
under $\alpha$; thus the central charges' eigenvalues should not
depend on $\kappa$ either. What has happened is that, by using
a different basis for $\g_R$, which allowed us to write the action
of the $\g_+$ charges in untwisted diagonal form, we have introduced
$\kappa$-dependence into the $L\circ R$ basis of the central charges.
That is, the price we have to pay for simplifying the action of $\alpha$
is that the central charges $\mathbb{C}$ and $\mathbb{C}^\dagger$
become dependent on $\kappa$. We resolve this by introducing complementary
central charges
\begin{align}
\overline{\mathbb{C}} & =\mathbb{C}\circ1-1\circ\mathbb{C}=ab-\tilde{a}\tilde{b}=(1+\kappa^{2})ab,\nonumber \\
\overline{\mathbb{C}}^{\dagger} & =\mathbb{C}^{\dagger}\circ1-1\circ\mathbb{C}^{\dagger}=cd-\tilde{c}\tilde{d}=(1+\kappa^{-2})ab,\nonumber \\
\overline{\mathbb{H}} & =\mathbb{H}\circ1-1\circ\mathbb{H}=ad+bc-\tilde{a}\tilde{d}-\tilde{b}\tilde{c}=0.\label{c_bar}
\end{align}
This formal enlargement of the algebra does not -- cannot -- add
any new constraints to the system. The charge $\overline{\mathbb{H}}$
has zero eigenvalues on the bulk
and boundary reps independently of $\kappa$ and we have introduced it
merely to enable us to write the Yangian charges in a nicely symmetric form.
The new charges $\overline{\mathbb{C}}$ and $\overline{\mathbb{C}}^{\dagger}$ will
have non-zero eigenvalues on the bulk reps when $\kappa^{2}=+1$,
while the charges $\mathbb{C}$ and $\mathbb{C}^{\dagger}$ then
vanish; and vice versa for the $\kappa^{2}=-1$ case. Thus there
are always exactly three non-trivial central charges in the system.

\paragraph{The Yangian algebra.} We now have
enough ingredients  to write down the general form of the boundary
Yangian. Following the discussion in section 2 we define
the achiral twisted boundary Yangian $\Y(\g_{L}\times\g_{R},\g_{+})$
to be generated by the Lie algebra generators (\ref{diag_symm}) and
(\ref{comp_diag_symm}), central charges and the twisted Yangian charges (\ref{Y_twist}).

\begin{align}
\widetilde{\mathbb{R}}_{\hat{a}}^{\enskip\hat{b}} & =\hat{\overline{\mathbb{R}}}_{\hat{a}}^{\enskip\hat{b}}-\frac{1}{4}\mathbb{R}_{\check{a}}^{\enskip\check{c}}\,\mathbb{\overline{R}}_{\check{c}}^{\enskip\check{b}}+\frac{1}{4}\mathbb{R}_{\check{c}}^{\enskip\check{b}}\,\mathbb{\overline{R}}_{\check{a}}^{\enskip\check{c}}+\frac{1}{4}\mathbb{G}_{\check{a}}^{\enskip\check{\gamma}}\,\mathbb{\overline{Q}}_{\check{\gamma}}^{\enskip\check{b}}+\frac{1}{4}\mathbb{Q}_{\check{\gamma}}^{\enskip\check{b}}\,\mathbb{\overline{G}}_{\check{a}}^{\enskip\check{\gamma}}\nonumber \\
 & \qquad\quad-\frac{1}{8}\delta_{\check{a}}^{\check{b}}\,\mathbb{G}_{\check{c}}^{\enskip\check{\gamma}}\,\mathbb{\overline{Q}}_{\check{\gamma}}^{\enskip\check{c}}-\frac{1}{8}\delta_{\check{a}}^{\check{b}}\,\mathbb{Q}_{\check{\gamma}}^{\enskip\check{c}}\,\mathbb{\overline{G}}_{\check{c}}^{\enskip\check{\gamma}},\nonumber \\
\widetilde{\mathbb{L}}_{\hat{\alpha}}^{\enskip\hat{\beta}} & =\hat{\overline{\mathbb{L}}}_{\hat{\alpha}}^{\enskip\hat{\beta}}+\frac{1}{4}\mathbb{L}_{\check{\alpha}}^{\enskip\check{\gamma}}\,\mathbb{\overline{L}}_{\check{\gamma}}^{\enskip\check{\beta}}-\frac{1}{4}\mathbb{L}_{\check{\gamma}}^{\enskip\check{\beta}}\,\mathbb{\overline{L}}_{\check{\alpha}}^{\enskip\check{\gamma}}-\frac{1}{4}\mathbb{G}_{\check{c}}^{\enskip\check{\beta}}\,\mathbb{\overline{Q}}_{\check{\alpha}}^{\enskip\check{c}}-\frac{1}{4}\mathbb{Q}_{\check{\alpha}}^{\enskip\check{c}}\,\mathbb{\overline{G}}_{\check{c}}^{\enskip\check{\beta}}\nonumber \\
 & \qquad\quad+\frac{1}{8}\delta_{\check{\alpha}}^{\check{\beta}}\,\mathbb{G}_{\check{c}}^{\enskip\check{\gamma}}\,\mathbb{\overline{Q}}_{\check{\gamma}}^{\enskip\check{c}}+\frac{1}{8}\delta_{\check{\alpha}}^{\check{\beta}}\,\mathbb{Q}_{\check{\gamma}}^{\enskip\check{c}}\,\mathbb{\overline{G}}_{\check{c}}^{\enskip\check{\gamma}},\nonumber\\
\widetilde{\mathbb{Q}}_{\hat{\alpha}}^{\enskip\hat{a}} & =\hat{\overline{\mathbb{Q}}}_{\hat{\alpha}}^{\enskip\hat{a}}-\frac{1}{4}\mathbb{Q}_{\check{\alpha}}^{\enskip\check{c}}\,\mathbb{\overline{R}}_{\check{c}}^{\enskip\check{a}}+\frac{1}{4}\mathbb{R}_{\check{c}}^{\enskip\check{a}}\,\mathbb{\overline{Q}}_{\check{\alpha}}^{\enskip\check{c}}-\frac{1}{4}\mathbb{Q}_{\check{\gamma}}^{\enskip\check{a}}\,\mathbb{\overline{L}}_{\check{\alpha}}^{\enskip\check{\gamma}}+\frac{1}{4}\mathbb{L}_{\check{\alpha}}^{\enskip\check{\gamma}}\,\mathbb{\overline{Q}}_{\check{\gamma}}^{\enskip\check{a}}\nonumber \\
 & \qquad\quad\,+\frac{1}{8}\mathbb{Q}_{\check{\alpha}}^{\enskip\check{a}}\,\overline{\mathbb{H}}+\frac{1}{8}\mathbb{H}\,\mathbb{\overline{Q}}_{\check{\alpha}}^{\enskip\check{a}}-\frac{1}{4}\varepsilon_{\check{\alpha}\check{\gamma}}\,\varepsilon^{\check{a}\check{d}}\,\mathbb{C}\,\mathbb{\overline{G}}_{\check{d}}^{\enskip\check{\gamma}}+\frac{1}{4}\varepsilon_{\check{\alpha}\check{\gamma}}\,\varepsilon^{\check{a}\check{d}}\,\mathbb{G}_{\check{d}}^{\enskip\check{\gamma}}\,\overline{\mathbb{C}},\nonumber \\
\widetilde{\mathbb{G}}_{\hat{a}}^{\enskip\hat{\alpha}} & =\hat{\overline{\mathbb{G}}}_{\hat{a}}^{\enskip\hat{\alpha}}+\frac{1}{4}\mathbb{G}_{\check{c}}^{\enskip\check{\alpha}}\,\mathbb{\overline{R}}_{\check{a}}^{\enskip\check{c}}-\frac{1}{4}\mathbb{R}_{\check{a}}^{\enskip\check{c}}\,\mathbb{\overline{G}}_{\check{c}}^{\enskip\check{\alpha}}+\frac{1}{4}\mathbb{G}_{\check{a}}^{\enskip\check{\gamma}}\,\mathbb{\overline{L}}_{\check{\gamma}}^{\enskip\check{\alpha}}-\frac{1}{4}\mathbb{L}_{\check{\gamma}}^{\enskip\check{\alpha}}\,\mathbb{\overline{G}}_{\check{a}}^{\enskip\check{\gamma}}\nonumber \\
 & \qquad\quad\,+\frac{1}{8}\mathbb{G}_{\check{a}}^{\enskip\check{\alpha}}\,\overline{\mathbb{H}}-\frac{1}{8}\mathbb{H}\,\mathbb{\overline{G}}_{\check{a}}^{\enskip\check{\alpha}}+\frac{1}{4}\varepsilon_{\check{a}\check{c}}\,\varepsilon^{\check{\alpha}\check{\gamma}}\,\mathbb{C}^{\dagger}\,\mathbb{\overline{Q}}_{\check{\gamma}}^{\enskip\check{c}}-\frac{1}{4}\varepsilon_{\check{a}\check{c}}\,\varepsilon^{\check{\alpha}\check{\gamma}}\,\mathbb{Q}_{\check{\gamma}}^{\enskip\check{c}}\,\mathbb{\overline{C}}^{\dagger},\nonumber\\
\widetilde{\mathbb{C}} & =\hat{\overline{\mathbb{C}}}+\frac{1}{4}\mathbb{H}\,\overline{\mathbb{C}}-\frac{1}{4}\mathbb{C}\,\overline{\mathbb{H}},\nonumber\\
\widetilde{\mathbb{C}}^{\dagger} &=\hat{\overline{\mathbb{C}}}^{\dagger}-\frac{1}{4}\mathbb{H}\,\overline{\mathbb{C}}^{\dagger}+\frac{1}{4}\mathbb{C}^{\dagger}\,\overline{\mathbb{H}}.\label{Y_twist}
\end{align}
Here the hat-bar operators $\hat{\overline{\mathbb{J}}}$ are the
grade-1 partners of the complementary charges (\ref{comp_diag_symm})
and (\ref{c_bar}). This is the general form of the twisted boundary
Yangian for the reflection from the $D5$-brane
and represents the explicit realization of \eqref{twist2}.
The co-products of the twisted charges (\ref{Y_twist}) have the
canonical form
\begin{align}
\Delta\widetilde{\mathbb{R}}_{\check{a}}^{\enskip\check{b}} & =\widetilde{\mathbb{R}}_{\check{a}}^{\enskip\check{b}}\otimes1+1\otimes\widetilde{\mathbb{R}}_{\check{a}}^{\enskip\check{b}}+\frac{1}{2}\overline{\mathbb{R}}_{\check{a}}^{\enskip\check{c}}\otimes\mathbb{R}_{\check{c}}^{\enskip\check{b}}-\frac{1}{2}\overline{\mathbb{R}}_{\check{c}}^{\enskip\check{b}}\otimes\mathbb{R}_{\check{a}}^{\enskip\check{c}}+\frac{1}{2}\overline{\mathbb{G}}_{\check{a}}^{\enskip\check{\gamma}}\otimes\mathbb{Q}_{\check{\gamma}}^{\enskip\check{b}}+\frac{1}{4}\overline{\mathbb{Q}}_{\check{\gamma}}^{\enskip\check{b}}\otimes\mathbb{G}_{\check{a}}^{\enskip\check{\gamma}}\nonumber\\
 & \qquad-\frac{1}{4}\delta_{\check{a}}^{\check{b}}\,\overline{\mathbb{G}}_{\check{c}}^{\enskip\check{\gamma}}\otimes\mathbb{Q}_{\check{\gamma}}^{\enskip\check{c}}-\frac{1}{4}\delta_{\check{a}}^{\check{b}}\,\overline{\mathbb{Q}}_{\check{\gamma}}^{\enskip\check{c}}\otimes\mathbb{G}_{\check{c}}^{\enskip\check{\gamma}},\nonumber\\
\Delta\widetilde{\mathbb{L}}_{\check{\alpha}}^{\enskip\check{\beta}} & =\widetilde{\mathbb{L}}_{\check{\alpha}}^{\enskip\check{\beta}}\otimes1+1\otimes\widetilde{\mathbb{L}}_{\check{\alpha}}^{\enskip\check{\beta}}+\frac{1}{2}\overline{\mathbb{L}}_{\check{\alpha}}^{\enskip\check{\gamma}}\otimes\mathbb{L}_{\check{\gamma}}^{\enskip\check{\beta}}-\frac{1}{2}\overline{\mathbb{L}}_{\check{\gamma}}^{\enskip\check{\beta}}\otimes\mathbb{L}_{\check{\alpha}}^{\enskip\check{\gamma}}-\frac{1}{2}\overline{\mathbb{G}}_{\check{c}}^{\enskip\check{\beta}}\otimes\mathbb{Q}_{\check{\alpha}}^{\enskip\check{c}}-\frac{1}{2}\overline{\mathbb{Q}}_{\check{\alpha}}^{\enskip\check{c}}\otimes\mathbb{G}_{\check{c}}^{\enskip\check{\beta}}\nonumber\\
 & \qquad+\frac{1}{4}\delta_{\check{\alpha}}^{\check{\beta}}\,\overline{\mathbb{G}}_{\check{c}}^{\enskip\check{\gamma}}\otimes\mathbb{Q}_{\check{\gamma}}^{\enskip\check{c}}+\frac{1}{4}\delta_{\check{\alpha}}^{\check{\beta}}\,\overline{\mathbb{Q}}_{\check{\gamma}}^{\enskip\check{c}}\otimes\mathbb{G}_{\check{c}}^{\enskip\check{\gamma}},\nonumber\\
\Delta\widetilde{\mathbb{Q}}_{\check{\alpha}}^{\enskip\check{a}} & =\widetilde{\mathbb{Q}}_{\check{\alpha}}^{\enskip\check{a}}\otimes1+1\otimes\widetilde{\mathbb{Q}}_{\check{\alpha}}^{\enskip\check{a}}-\frac{1}{2}\overline{\mathbb{Q}}_{\check{\alpha}}^{\enskip\check{c}}\otimes\mathbb{R}_{\check{c}}^{\enskip\check{a}}+\frac{1}{2}\overline{\mathbb{R}}_{\check{c}}^{\enskip\check{a}}\otimes\mathbb{Q}_{\check{\alpha}}^{\enskip\check{c}}-\frac{1}{2}\overline{\mathbb{Q}}_{\check{\gamma}}^{\enskip\check{a}}\otimes\mathbb{L}_{\check{\alpha}}^{\enskip\check{\gamma}}+\frac{1}{2}\overline{\mathbb{L}}_{\check{\alpha}}^{\enskip\check{\gamma}}\otimes\mathbb{Q}_{\check{\gamma}}^{\enskip\check{a}}\nonumber\\
 & \qquad+\frac{1}{4}\overline{\mathbb{H}}\otimes\mathbb{Q}_{\check{\alpha}}^{\enskip\check{a}}-\frac{1}{4}\overline{\mathbb{Q}}_{\check{\alpha}}^{\enskip\check{a}}\otimes\mathbb{H}-\frac{1}{2}\varepsilon_{\check{\alpha}\check{\gamma}}\,\varepsilon^{\check{a}\check{d}}\,\overline{\mathbb{C}}\otimes\mathbb{G}_{\check{d}}^{\enskip\check{\gamma}}+\frac{1}{2}\varepsilon_{\check{\alpha}\check{\gamma}}\,\varepsilon^{\check{a}\check{d}}\,\overline{\mathbb{G}}_{\check{d}}^{\enskip\check{\gamma}}\otimes\mathbb{C},\nonumber\\
\Delta\widetilde{\mathbb{G}}_{\check{a}}^{\enskip\check{\alpha}} & =\widetilde{\mathbb{G}}_{\check{a}}^{\enskip\check{\alpha}}\otimes1+1\otimes\widetilde{\mathbb{G}}_{\check{a}}^{\enskip\check{\alpha}}+\frac{1}{2}\overline{\mathbb{G}}_{\check{c}}^{\enskip\check{\alpha}}\otimes\mathbb{R}_{\check{a}}^{\enskip\check{c}}-\frac{1}{2}\overline{\mathbb{R}}_{\check{a}}^{\enskip\check{c}}\otimes\mathbb{G}_{\check{c}}^{\enskip\check{\alpha}}+\frac{1}{2}\overline{\mathbb{G}}_{\check{a}}^{\enskip\check{\gamma}}\otimes\mathbb{L}_{\check{\gamma}}^{\enskip\check{\alpha}}-\frac{1}{2}\overline{\mathbb{L}}_{\check{\gamma}}^{\enskip\check{\alpha}}\otimes\mathbb{G}_{\check{a}}^{\enskip\check{\gamma}}\nonumber\\
 & \qquad-\frac{1}{4}\overline{\mathbb{H}}\otimes\mathbb{G}_{\check{a}}^{\enskip\check{\alpha}}+\frac{1}{4}\overline{\mathbb{G}}_{\check{a}}^{\enskip\check{\alpha}}\otimes\mathbb{H}+\frac{1}{2}\varepsilon_{\check{a}\check{c}}\,\varepsilon^{\check{\alpha}\check{\gamma}}\,\overline{\mathbb{C}}^{\dagger}\otimes\mathbb{Q}_{\check{\gamma}}^{\enskip\check{c}}-\frac{1}{2}\varepsilon_{\check{a}\check{c}}\,\varepsilon^{\check{\alpha}\check{\gamma}}\,\overline{\mathbb{Q}}_{\check{\gamma}}^{\enskip\check{c}}\otimes\mathbb{C}^{\dagger},\nonumber\\
\Delta\widetilde{\mathbb{C}} & =\widetilde{\mathbb{C}}\otimes1+1\otimes\widetilde{\mathbb{C}}+\frac{1}{2}\overline{\mathbb{C}}\otimes\mathbb{H}-\frac{1}{2}\overline{\mathbb{H}}\otimes\mathbb{C},\nonumber\\
\Delta\widetilde{\mathbb{C}}^{\dagger} & =\widetilde{\mathbb{C}}^{\dagger}\otimes1+1\otimes\widetilde{\mathbb{C}}^{\dagger}-\frac{1}{2}\overline{\mathbb{C}}^{\dagger}\otimes\mathbb{H}+\frac{1}{2}\overline{\mathbb{H}}\otimes\mathbb{C}^{\dagger},\label{Y(gxg,g)}
\end{align}
as expected from \eqref{coideal2}. Note that the terms of the form 
$1\otimes\widetilde{\mathbb{J}}$
annihilate the boundary and give no contribution to the explicit calculations.
Also note that the expressions above may be reduced to a more compact
and transparent form using the Lie algebra relations (\ref{psu(2|2)_algebra}).
We will do this by considering the reflection from the vertical and
horizontal $D5$-branes separately.


\subsection{The horizontal D5-brane}

The boundary algebra of the horizontal $D5$-brane is acquired from (\ref{diag_symm})
by setting $\kappa=-1$. As was shown in \cite{CY}, the boundary
carries no degrees of freedom in the scattering theory, and thus
is a singlet of $\mathfrak{psl}(2|2)_{+}$. The total central charges
$\mathbb{C}$ and $\mathbb{C}^{\dagger}$  vanish with respect
to the boundary symmetry,
\begin{equation}
\left\langle 0,0;H,C,C^{\dagger}\right\rangle \circ\left\langle 0,0;H,-C,-C^{\dagger}\right\rangle =\left\{ 0,0,2H,0,0\right\},
\end{equation}
and the bulk magnon transforms in the tensor representation
\begin{equation}
\boxslash_{(a,b,c,d)} \circ \; \widetilde{\boxslash}_{(a,-b,-c,d)},\label{(p,-p) rep}
\end{equation}
with respect to the boundary algebra. The reflection matrix is
simply a map
\begin{equation}
\mathcal{K}^{h}:\boxslash\circ\widetilde{\boxslash}\otimes1\rightarrow\boxslash\circ\widetilde{\boxslash}\otimes1,
\end{equation}
and may be neatly represented on  superspace as an
operator
\begin{equation}
\mathcal{K}^{h}:\quad\mathcal{V}\left(p,\zeta\right)\circ\mathcal{V}(-p,\zeta e^{ip})\rightarrow\mathcal{V}(-p,\zeta)\circ\mathcal{V}(p,\zeta e^{-ip}),
\end{equation}
where $\mathcal{V}(p,\zeta)$ is the corresponding vector space.
Thus $\mathcal{K}^{h}$ differs from the bulk $S$-matrix
$\mathcal{S}(p,-p)$ by an overall phase at most.

The convenient parametrization of the representation labels of
a single magnon in the bulk is
\begin{equation}
a=\sqrt{\frac{g}{2}}\eta,\quad b=\sqrt{\frac{g}{2}}\frac{i\zeta}{\eta}\left(\frac{x^{+}}{x^{-}}-1\right),\quad
 c=-\sqrt{\frac{g}{2}}\frac{\eta}{\zeta x^{+}},\quad d=-\sqrt{\frac{g}{2}}\frac{x^{+}}{i\eta}\left(\frac{x^{-}}{x^{+}}-1\right),\label{abcd}
\end{equation}
where $\zeta=e^{2i\xi}$ is the magnon phase and unitarity implies
$\eta=e^{i\xi}e^{i\frac{\varphi}{2}}\sqrt{i\left(x^{-}-x^{+}\right)}$;
here the arbitrary phase factor $e^{i\varphi}$ parametrizes the freedom
in choosing $x^{\pm}$ and the standard choice is $\varphi=p/2$.

\begin{figure}
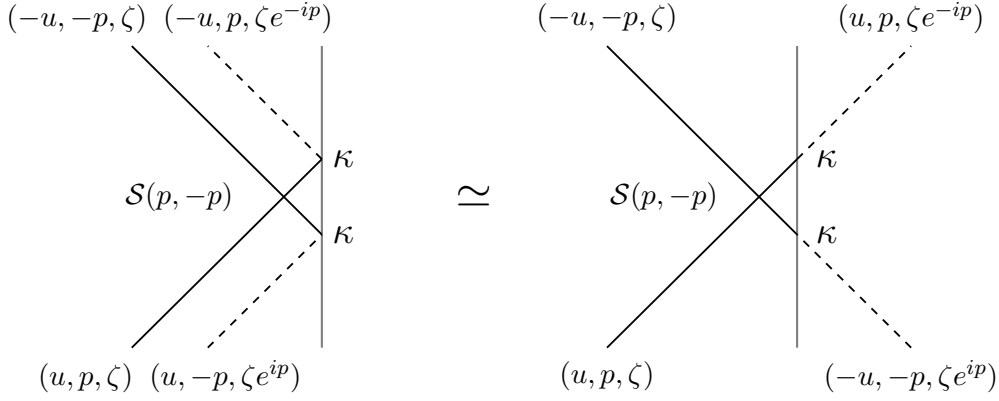

\begin{centering}
\btp
   \draw[gray,thick,-] (10,0) -- (10,16);
   \draw[-] (0,0) -- (10,10);
   \draw[-] (10,6) -- (0,16);
   \draw[-,dashed] (4,0) -- (10,6) node [right] {$\Kappa$};
   \draw[-,dashed] (10,10) node [right] {$\Kappa$} -- (4,16);
   \draw (2.5,8) node {$\mathcal{S}(p,-p)$}  (2,-1.5) node {$(u,p,\zeta)\;\;(u,-p,\zeta e^{ip})$}
   (2,17.5) node {$(-u,-p,\zeta)\;\;(-u,p,\zeta e^{-ip})$};
   \draw[gray,thick,-] (35,0) -- (35,16);
   \draw[-] (25,0) -- (35,10);
   \draw[-,dashed] (35,10) node [right] {$\;\Kappa$} -- (41,16);
   \draw[-,dashed] (41,0) -- (35,6) node [right] {$\;\Kappa$};
   \draw[-] (35,6) -- (25,16);
   \draw (18,8) node {\text{\LARGE$\simeq$}} (28,8) node {$\mathcal{S}(p,-p)$} (25,-1.5) node {$(u,p,\zeta)$}
   (41,-1.5) node {$(-u,-p,\zeta e^{ip})$}  (25,17.5) node {$(-u,-p,\zeta)$} (41,17.5) node {$(u,p,\zeta e^{-ip})$};
\etp
\par\end{centering}
\hspace{0.5in}%
\parbox[c]{5in}{%
\caption{Unfolding of the reflection from the horizontal $D5$-brane. Solid
lines correspond to the left reps while the dotted lines correspond
to right reps. The vertical gray lines correspond to the singlet boundary
which acts merely as an achiral map $\Kappa$ mapping right (left) reps into
left (right) reps (and conjugates multiplets by maping $u\mapsto-u$ in the unfolded picture).
The left and right sides of the figure are related through the conjugation map T.}
\label{fig_Kh}}
\end{figure}

The boundary is achiral in the sense that the incoming $L$ state
becomes a $R$ state after reflection and vice versa. This feature
of the achiral boundary may be neatly displayed graphically (see figure
\ref{fig_Kh}, left side). The picture of the reflection nicely accommodates the fact
that $\mathcal{K}^{h}$ is equivalent to $\mathcal{S}$ as discussed
above and suggests that it should be related as
\begin{equation}
\mathcal{K}^{h}=\Kappa \cdot \mathcal{S}(p,-p)\cdot \Kappa\,,
\end{equation}
with $\Kappa\;$ here\footnote{This enlarged $\Kappa$ denoting the achiral map
should not be confused with $\kappa$ in (\ref{diag_symm}).} being the achiral map 
$\Kappa:1\circ\widetilde{\boxslash}\otimes1 \rightarrow \boxslash\circ1\otimes1$
for an incoming right state and 
$\Kappa: \boxslash\circ1\otimes1 \rightarrow 1\circ\widetilde{\boxslash}\otimes1$
for an incoming left state. This relation implies that the structure of the boundary
Yangian for the horizontal D5-brane should be related to the bulk Yangian (\ref{Y(g)}) by
(\ref{YYre}). Let us show this explicitly.

The boundary is a singlet; thus the only surviving terms in (\ref{Y(gxg,g)})
are of the form $\widetilde{\mathbb{J}}\otimes1$, since all non-local two-site
operators of the form $\mathbb{J}\otimes\mathbb{J}$ annihilate the
boundary and give no contribution.\footnote{See e.g.\ \cite{GZ94} for
the formulation of the scattering theory on the half-line.} Then using
(\ref{Y_twist},\,\ref{diag_symm},\,\ref{comp_diag_symm},\,\ref{c_bar})
and performing some Lie algebra manipulations, (\ref{Y(gxg,g)}) become
\begin{align*}
\Delta\widetilde{\mathbb{R}}_{\check{a}}^{\enskip\check{b}} & =\Bigl(\hat{\mathbb{R}}_{\check{a}}^{\enskip\check{b}}\circ1-1\circ\hat{\mathbb{R}}_{\check{a}}^{\enskip\check{b}}+\frac{1}{2}\mathbb{R}_{\check{a}}^{\enskip\check{c}}\circ\mathbb{R}_{\check{c}}^{\enskip\check{b}}-\frac{1}{2}\mathbb{R}_{\check{c}}^{\enskip\check{b}}\circ\mathbb{R}_{\check{a}}^{\enskip\check{c}}-\frac{1}{2}\mathbb{G}_{\check{a}}^{\enskip\check{\gamma}}\circ\mathbb{Q}_{\check{\gamma}}^{\enskip\check{b}}-\frac{1}{2}\mathbb{Q}_{\check{\gamma}}^{\enskip\check{b}}\circ\mathbb{G}_{\check{a}}^{\enskip\check{\gamma}}\nonumber\\
 & \qquad+\frac{1}{4}\delta_{\check{a}}^{\check{b}}\,\mathbb{G}_{\check{c}}^{\enskip\check{\gamma}}\circ\mathbb{Q}_{\check{\gamma}}^{\enskip\check{c}}+\frac{1}{4}\delta_{\check{a}}^{b}\,\mathbb{Q}_{\check{\gamma}}^{\enskip\check{c}}\circ\mathbb{G}_{\check{c}}^{\enskip\check{\gamma}}\Bigr)\otimes1,\nonumber\\
\Delta\widetilde{\mathbb{L}}_{\check{\alpha}}^{\enskip\check{\beta}} & =\Bigl(\hat{\mathbb{L}}_{\check{\alpha}}^{\enskip\check{\beta}}\circ1-1\circ\hat{\mathbb{L}}_{\check{\alpha}}^{\enskip\check{\beta}}-\frac{1}{2}\mathbb{L}_{\check{\alpha}}^{\enskip\check{\gamma}}\circ\mathbb{L}_{\check{\gamma}}^{\enskip\check{\beta}}+\frac{1}{2}\mathbb{L}_{\check{\gamma}}^{\enskip\check{\beta}}\circ\mathbb{L}_{\check{\alpha}}^{\enskip\check{\gamma}}+\frac{1}{2}\mathbb{G}_{\check{c}}^{\enskip\check{\beta}}\circ\mathbb{Q}_{\check{\alpha}}^{\enskip\check{c}}+\frac{1}{2}\mathbb{Q}_{\check{\alpha}}^{\enskip\check{c}}\circ\mathbb{G}_{\check{c}}^{\enskip\check{\beta}}\nonumber\\
 & \qquad-\frac{1}{4}\delta_{\check{\alpha}}^{\check{\beta}}\,\mathbb{G}_{\check{c}}^{\enskip\check{\gamma}}\circ\mathbb{Q}_{\check{\gamma}}^{\enskip\check{c}}-\frac{1}{4}\delta_{\check{\alpha}}^{\check{\beta}}\,\mathbb{Q}_{\check{\gamma}}^{\enskip\check{c}}\circ\mathbb{G}_{\check{c}}^{\enskip\check{\gamma}}\Bigr)\otimes1,\nonumber\\
\Delta\widetilde{\mathbb{Q}}_{\check{\alpha}}^{\enskip\check{a}} & =\Bigl(\hat{\mathbb{Q}}_{\check{\alpha}}^{\enskip\check{a}}\circ1-1\circ\hat{\mathbb{Q}}_{\check{\alpha}}^{\enskip\check{a}}+\frac{1}{2}\mathbb{Q}_{\check{\alpha}}^{\enskip\check{c}}\circ\mathbb{R}_{\check{c}}^{\enskip\check{a}}-\frac{1}{2}\mathbb{R}_{\check{c}}^{\enskip\check{a}}\circ\mathbb{Q}_{\check{\alpha}}^{\enskip\check{c}}+\frac{1}{2}\mathbb{Q}_{\check{\gamma}}^{\enskip\check{a}}\circ\mathbb{L}_{\check{\alpha}}^{\enskip\check{\gamma}}-\frac{1}{2}\mathbb{L}_{\check{\alpha}}^{\enskip\check{\gamma}}\circ\mathbb{Q}_{\check{\gamma}}^{\enskip\check{a}}\nonumber\\
 & \qquad+\frac{1}{4}\mathbb{Q}_{\check{\alpha}}^{\enskip\check{a}}\circ\mathbb{H}-\frac{1}{4}\mathbb{H}\circ\mathbb{Q}_{\check{\alpha}}^{\enskip\check{a}}+\frac{1}{2}\varepsilon_{\check{\alpha}\check{\gamma}}\,\varepsilon^{ad}\,\mathbb{C}\circ\mathbb{G}_{d}^{\enskip\check{\gamma}}-\frac{1}{2}\varepsilon_{\check{\alpha}\check{\gamma}}\,\varepsilon^{ad}\,\mathbb{G}_{\check{d}}^{\enskip\check{\gamma}}\circ\mathbb{C}\Bigr)\otimes1,\nonumber\\
\Delta\widetilde{\mathbb{G}}_{\check{a}}^{\enskip\check{\alpha}} & =\Bigl(\hat{\mathbb{G}}_{\check{a}}^{\enskip\check{\alpha}}\circ1-1\circ\hat{\mathbb{G}}_{\check{a}}^{\enskip\check{\alpha}}-\frac{1}{2}\mathbb{G}_{\check{c}}^{\enskip\check{\alpha}}\circ\mathbb{R}_{\check{a}}^{\enskip\check{c}}+\frac{1}{2}\mathbb{R}_{\check{a}}^{\enskip\check{c}}\circ\mathbb{G}_{\check{c}}^{\enskip\check{\alpha}}-\frac{1}{2}\mathbb{G}_{\check{a}}^{\enskip\check{\gamma}}\circ\mathbb{L}_{\check{\gamma}}^{\enskip\check{\alpha}}+\frac{1}{2}\mathbb{L}_{\check{\gamma}}^{\enskip\check{\alpha}}\circ\mathbb{G}_{\check{a}}^{\enskip\check{\gamma}}\nonumber\\
 & \qquad-\frac{1}{4}\mathbb{G}_{\check{a}}^{\enskip\check{\alpha}}\circ\mathbb{H}+\frac{1}{4}\mathbb{H}\circ\mathbb{G}_{\check{a}}^{\enskip\check{\alpha}}-\frac{1}{2}\varepsilon_{\check{a}\check{c}}\,\varepsilon^{\check{\alpha}\check{\gamma}}\,\mathbb{C}^{\dagger}\circ\mathbb{Q}_{\check{\gamma}}^{\enskip\check{c}}+\frac{1}{2}\varepsilon_{\check{a}\check{c}}\,\varepsilon^{\check{\alpha}\check{\gamma}}\,\mathbb{Q}_{\check{\gamma}}^{\enskip\check{c}}\circ\mathbb{C}^{\dagger}\Bigr)\otimes1,\nonumber
\end{align*}
\begin{align}
\Delta\widetilde{\mathbb{C}} & =\Bigl(\hat{\mathbb{C}}\circ1-1\circ\hat{\mathbb{C}}-\frac{1}{2}\mathbb{H}\circ\mathbb{C}+\frac{1}{2}\mathbb{C}\circ\mathbb{H}\Bigr)\otimes1,\nonumber \\
\Delta\widetilde{\mathbb{C}}^{\dagger} & =\Bigl(\hat{\mathbb{C}}^{\dagger}\circ1-1\circ\mathbb{\hat{C}}^{\dagger}+\frac{1}{2}\mathbb{H}\circ\mathbb{C}^{\dagger}-\frac{1}{2}\mathbb{C}^{\dagger}\circ\mathbb{H}\Bigr)\otimes1.
\label{Y_cop_D5h}
\end{align}
We have checked that these co-products commute with the reflection
matrix $\mathcal{K}^{h}$ calculated in \cite{CRY} and also with
the two-magnon bound-state reflection matrix which is constructed
from $S^{BB}(p_{1},p_{2})$ \cite{MR1} by setting $p_{2}:=-p_{1}$.

It is easy to observe that these co-products have almost the same
form as (\ref{Y(g)}), as we expected. The crucial difference is the
negative sign of terms of the form $1\circ\hat{\mathbb{J}}$ in (\ref{Y_cop_D5h}),
in contrast to (\ref{Y(g)}). This is the outcome of the graded map
$1\circ(-1)^l$ relating $\tilde\Delta$ to the usual $\Delta$ in (\ref{YYre}).
In this particular case it has a lucid physical interpretation.
Consider the scattering in the bulk of two magnons with momenta $p$ and
$-p$. The residual symmetry of such scattering is described by (\ref{Y(g)}).
The rapidities of the states in the bulk are $u$ and $-u$ and are
facing the same direction as their momenta. But in the case of a single
bulk magnon reflecting from the horizontal $D5$-brane the rapidity
of the right rep, which has the effective momentum $-p$ with respect
to the boundary algebra, is $u$ and faces the physical direction,
but not the effective one i.e. is not $-u$). Thus this minus sign
difference is explicitly seen in the co-products (\ref{Y_cop_D5h}).

Interestingly, in the unfolded picture of the reflection (the right side of
figure \ref{fig_Kh}), which is related to the left side by the map T (\ref{T}),
the rapidity of the right rep is facing the same direction as the momentum.
This is because the map T not only re-orders the states, but also sends
$V_R \mapsto \bar{V}_R$, $u \mapsto -u$. Thus the
reflection from the boundary in the unfolded picture,
\begin{equation}
\mathcal{K}^{h}:\boxslash\otimes1\otimes\widetilde{\boxslash}
  \rightarrow \boxslash\otimes1\otimes\widetilde{\boxslash}, \label{Kh_unf}
\end{equation}
may be regarded as a `scattering through the boundary' and is governed by the Yangian
\begin{equation}
 \Delta \tilde{\mathbb{J}}^A=\hat{\mathbb{J}}^A\otimes1\otimes1+1\otimes1\otimes\hat{\mathbb{J}}^A
  +\frac{1}{2}f^A_{\;\,BC}\,\mathbb{J}^B\otimes1\otimes\mathbb{J}^C, \label{Y_Kh_unf}
\end{equation}
which is equivalent to (\ref{Y(g)}) (by removing the middle singlet in (\ref{Kh_unf}) and (\ref{Y_Kh_unf})
as it effectively plays no role).


\subsection{The vertical D5-brane}

The boundary algebra of the vertical $D5$-brane is acquired from (\ref{diag_symm}) by setting
$\kappa=-i$. We will consider reflection from the right boundary,
which carries a $\check{\boxslash}$ spanned by the fields $\phi^{a}$ and
$\psi^{\check{\alpha}\dot{1}}$ \cite{CY,CRY}. The scattering problem
under consideration is described by the triple tensor product
\begin{equation}
\boxslash_{(a,b,c,d)} \circ \; \widetilde{\boxslash}_{(a,b,c,d)} \otimes \check{\boxslash}_{(a_{B},b_{B},c_{B},d_{B})},\label{LRB}
\end{equation}
where once again $\circ$ describes $L\circ R$ reps of the bulk magnon while $\otimes$
describes the usual tensor product of $bulk \otimes boundary$ reps and the labels of the boundary rep are
\begin{equation}
a_{B} = \sqrt{g}\eta_{B},\quad b_{B} = -\sqrt{g}\frac{i\zeta}{\eta_{B}},\quad
 c_{B} = -\sqrt{g}\frac{\eta_{B}}{\zeta x_{B}},\quad d_{B} = \sqrt{g}\frac{x_{B}}{i\eta_{B}}.\label{abcd_B}
\end{equation}

\begin{figure}[h]
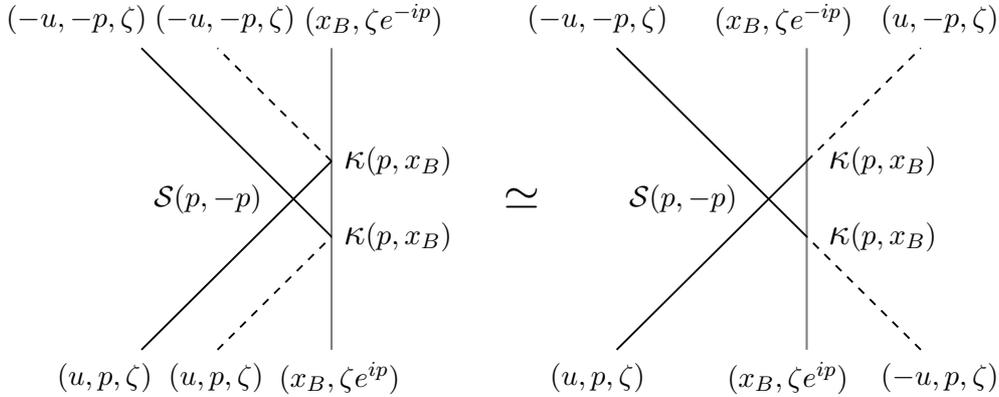

\begin{centering}
\btp
   \draw[gray,thick,-] (10,0) -- (10,16);
   \draw[-] (0,0) -- (10,10);
   \draw[-] (10,6) -- (0,16);
   \draw[-,dashed] (4,0) -- (10,6);
   \draw[-,dashed] (10,10) -- (4,16);
   \draw (3.5,8) node {$\mathcal{S}(p,-p)$} (13.5,6) node {$\Kappa(p,x_B)$} (13.5,10) node {$\Kappa(p,x_B)$}
         (1,-1.5) node {$(u,p,\zeta)\;\;(u,p,\zeta)$} (10,-1.5) node {$\;\;(x_B,\zeta e^{ip})$}
         (0.5,17.5) node {$(-u,-p,\zeta)\;(-u,-p,\zeta)$} (12,17.5) node {$\;(x_B,\zeta e^{-ip})$};

   \draw[gray,thick,-] (35,0) -- (35,16);
   \draw[-] (25,0) -- (35,10);
   \draw[-,dashed] (35,10) -- (41,16);
   \draw[-,dashed] (41,0) -- (35,6);
   \draw[-] (35,6) -- (25,16);
   \draw (20,8) node {\text{\LARGE$\simeq$}}  (28.5,8) node {$\mathcal{S}(p,-p)$} (39,6) node {$\Kappa(p,x_B)$}
         (39,10) node {$\Kappa(p,x_B)$}  (24,-1.5) node {$(u,p,\zeta)$} (42,-1.5) node {$(-u,p,\zeta)$}
         (34,-1.5) node {$(x_B,\zeta e^{ip})$} (24,17.5) node {$(-u,-p,\zeta)$} (42,17.5) node {$(u,-p,\zeta)$} (34,17.5) node {$(x_B,\zeta e^{-ip})\;$};
\etp
\par\end{centering}
\hspace{0.5in}\parbox[c]{5in}{%
\caption{Unfolding of the reflection from the vertical $D5$-brane. Solid lines
correspond to the left reps while the dotted lines correspond to right
reps. The vertical gray lines correspond to the boundary rep. In the
contrast to the horizontal case, the achiral reflection not only maps
left (right) reps to right (left) reps but is also an intertwining
matrix mapping momentum $p\mapsto-p$.}
\label{fig_Kv}}
\end{figure}

The total central charges of the bulk magnon with the respect to the
boundary algebra are
\begin{equation}
\mathbb{C} = 2C, \quad \mathbb{C}^{\dagger} = 2C^{\dagger}, \quad \mathbb{H} = 2H,
\end{equation}
and satisfy the multiplet splitting condition
\begin{equation}
\mathbb{H}^{2}-\mathbb{C}\mathbb{C}^{\dagger}=1,
\end{equation}
according to which
\begin{equation}
\left\{ 0,0;2H,2C,2C^{\dagger}\right\} =\left\langle 1,0;2H,2C,2C^{\dagger}\right\rangle \oplus\left\langle 0,1;2H,2C,2C^{\dagger}\right\rangle
  =\boxslash\negthickspace\!\boxslash\oplus\twobv.
\end{equation}
Two scattering channels were considered in \cite{CRY},
\begin{align}
 & \mathcal{K}^{Ba}:\boxslash\negthickspace\!\boxslash\otimes\,\check{\boxslash} \rightarrow \boxslash\negthickspace\!\boxslash\otimes\,\check{\boxslash},\label{SymRef}\\
 & \overline{\mathcal{K}}^{Ba}:\twobv\otimes\check{\boxslash} \rightarrow \twobv\otimes\check{\boxslash},\label{AsymRef}
\end{align}
with the complete reflection matrix being
\begin{equation}
\mathcal{K}^{v}=k_{0}\,\mathcal{K}^{Ba}+\overline{\mathcal{K}}^{Ba}.
\end{equation}
The symmetric and antisymmetric reflection matrices $\mathcal{K}^{Ba}$
and $\overline{\mathcal{K}}^{Ba}$ were found using only the Lie algebra,
while the ratio $k_{0}$ between the symmetric and anti-symmetric reflections
was determined from the boundary Yang-Baxter equation. As
was shown using the Bethe ansatz technique in \cite{CRY},
reflection from the vertical $D5$-brane is achiral. Hence it may
be represented by a diagram (figure \ref{fig_Kv}) very similar to
the one describing the reflection from the horizontal $D5$-brane
(figure \ref{fig_Kh}). %

Thus, as one can see from figure \ref{fig_Kv}, the reflection
factorizes as a composition of a bulk $S$-matrix and two achiral
reflection matrices $\Kappa$, with
\begin{equation}
\mathcal{K}^{v}(p,p)=\Kappa(p,x_{B})\,\mathcal{S}(p,-p)\,\Kappa(p,x_{B}).\label{Kv_factor}
\end{equation}
The achiral reflection matrix $\Kappa$ in the folded picture
maps incoming right states into outgoing left states as
\begin{equation}
\Kappa:1\circ\widetilde{\boxslash}\otimes\check{\boxslash} \mapsto \boxslash\circ1\otimes\check{\boxslash},\label{K_ach_R}
\end{equation}
and incoming left states into outgoing right ones as
\begin{equation}
\Kappa:\boxslash\circ1\otimes\check{\boxslash} \mapsto 1\circ\widetilde{\boxslash}\otimes\check{\boxslash}.\label{K_ach_L}
\end{equation}
These two expressions may be combined into one using vector space notation
\begin{equation}
\Kappa:\mathcal{V}_{L(R)}(p,\zeta)\otimes\mathcal{V}_{B}(x_{B},\zeta e^{ip})
 \mapsto \mathcal{V}_{R(L)}(-p,\zeta e^{ip})\otimes\mathcal{V}_{B}(x_{B},\zeta),\label{K_ach}
\end{equation}
and thereby may be defined on superspace in the usual way
\begin{equation}
\Kappa(p,x_{B})=\sum_{i=1}^{10}k_{i}(p,x_{B})\,\Lambda_{i},
\end{equation}
where $\Lambda_{i}$ are the $\mathfrak{su}(2)\times\mathfrak{su}(2)$
invariant differential operators (see \cite{CRY,AF}). Invariance under the
boundary algebra (\ref{diag_symm}) fixes $k_{i}$ up to an overall phase to be
\begin{align}
k_{1} &= -\frac{x_{B}-x^{-}}{x_{B}+x^{+}}\frac{\eta\eta_{B}}{\tilde{\eta}\tilde{\eta}_{B}},
  && k_{2} = \frac{5(x^{+}-x_{B})x^{-}-3((x^{-})^{2}-x_{B}x^{+})}{(x_{B}+x^{+})(x^{-}+x^{+})}\frac{\eta\eta_{B}}{\tilde{\eta}\tilde{\eta}_{B}},\nonumber\\
k_{3} &=1,
  && k_{4} = \frac{5(x^{-}+x_{B})x^{+}-3((x^{+})^{2}+x_{B}x^{-})}{(x_{B}+x^{+})(x^{-}+x^{+})},\nonumber \\
k_{5} & = -\frac{x_{B}-x^{+}}{x_{B}+x^{+}}\frac{\eta}{\tilde{\eta}},
  && k_{6} = \frac{x_{B}+x^{-}}{x_{B}+x^{+}}\frac{\eta_{B}}{\tilde{\eta}_{B}},\nonumber \\
k_{7} &= \frac{i\sqrt{2}\,\zeta\, x_B(x_{B}-x^{+})(x^{-}-x^{+})}{(x_{B}+x^{+})(1+x_B x^{-})\tilde{\eta}\tilde{\eta}_{B}},
  && k_{8} = \frac{i\sqrt{2}(x_{B}+x^{-})\eta\eta_{B}}{\zeta(x_{B}+x^{+})(1-x_B x^{-})},\nonumber \\
k_{9} & = \sqrt{2}\,\frac{x^{-}-x^{+}}{x_{B}+x^{+}}\frac{\eta_{B}}{\tilde{\eta}},
 && k_{10} = -\sqrt{2}\,\frac{x_{B}}{x_{B}+x^{+}}\frac{\eta}{\tilde{\eta}_{B}}.
\end{align}
We have checked explicitly that the factorization (\ref{Kv_factor})
is correct. It obeys the Yang-Baxter relation
and the reflection coefficients coincide with the
ones found in \cite{CRY}. For example, the reflection of the bulk
state $\phi_{1}\circ\tilde{\phi}_{1}$ from the boundary state $\check{\phi}_{1}$
gives a relation
\begin{equation}
k_{0}\, k_{1}^{\text{\tiny(S)}}(p_{1},\zeta;x_{B}) = k_{1}(p,\zeta e^{-ip};x_{B})\, a_{1}(p,-p,\zeta)\, k_{1}(p,\zeta;x_{B}),
\end{equation}
where $a_{1}$ and $k_{1}^{\text{\tiny(S)}}$ are the coefficients
of the fundamental $S$-matrix and reflection matrix $\mathcal{K}^{Ba}$
respectively; all of them are spelled out in the appendices of
\cite{CRY}.\footnote{The achiral reflection matrix is
equivalent to the S-matrix by identifying $x^\pm = \pm x_B$ up to
a graded permutation and an extra factor of $-i$ in $k_{3,4,6,8,10}$ due to
the change of the basis \eqref{new_basis} for the right rep. This map
identifies the boundary magnon with a bulk magnon of momentum $p=\pi$.}

\begin{figure}
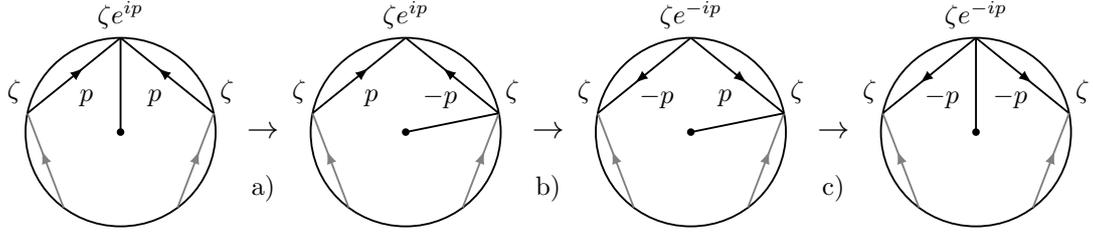

\begin{centering}
\btp\begin{scope}[decoration={markings, mark=at position 0.6 with {\arrow{latex}}}]{\small
   \draw (5,5) circle (5);   \draw[fill=black] (5,5) circle (1.5mm);
   \draw[gray,postaction={decorate}] (2,1) -- (0.1,6);
   \draw[postaction={decorate}] (0.1,6) node[above] {$\zeta\quad$} -- (5,10) node[above] {$\zeta e^{ip}$};
   \draw[postaction={decorate}] (9.9,6) node[above] {$\quad\zeta$} -- (5,10);
   \draw[gray,postaction={decorate}] (8,1) -- (9.9,6);
   \draw (5,5) -- (5,10);
   \draw (3.2,6.8) node {$p$} (6.8,6.8) node {$p$};

   \draw (20,5) circle (5);
   \draw[fill=black] (20,5) circle (1.5mm);
   \draw[gray,postaction={decorate}] (17,1) -- (15.1,6);
   \draw[postaction={decorate}] (15.1,6) node[above] {$\zeta\quad$} -- (20,10) node[above] {$\zeta e^{ip}$};
   \draw[postaction={decorate}] (24.9,6) node[above] {$\quad\zeta$} -- (20,10);
   \draw[gray,postaction={decorate}] (23,1) -- (24.9,6);
   \draw (20,5) -- (24.9,6);
   \draw (18.2,6.8) node {$p$} (21.8,6.8) node {$-p$};

   \draw (35,5) circle (5);
   \draw[fill=black] (35,5) circle (1.5mm);
   \draw[gray,postaction={decorate}] (32,1) -- (30.1,6);
   \draw[postaction={decorate}] (35,10) node[above] {$\zeta e^{-ip}$} -- (30.1,6) node[above] {$\zeta\quad$};
   \draw[postaction={decorate}] (35,10) -- (39.9,6) node[above] {$\quad\zeta$};
   \draw[gray,postaction={decorate}] (38,1) -- (39.9,6);
   \draw (35,5) -- (39.9,6);
   \draw (33.2,6.8) node {$-p$} (36.8,6.8) node {$p$};

   \draw (50,5) circle (5);
   \draw[fill=black] (50,5) circle (1.5mm);
   \draw[gray,postaction={decorate}] (47,1) -- (45.1,6);
   \draw[postaction={decorate}] (50,10) node[above] {$\zeta e^{-ip}$} -- (45.1,6) node[above] {$\zeta\quad$};
   \draw[postaction={decorate}] (50,10) -- (54.9,6) node[above] {$\quad\zeta$};
   \draw[gray,postaction={decorate}] (53,1) -- (54.9,6);
   \draw (50,5) -- (50,10);   \draw (48.2,6.8) node {$-p$} (51.8,6.8) node {$-p$};
   \draw(12.5,5) node {\large $\rightarrow$} (27.5,5) node {\large $\rightarrow$}(42.5,5) node {\large $\rightarrow$};
   \draw(12.5,2) node {a)} (27.5,2) node {b)} (42.5,2) node {c)};
}\end{scope}\etp
\caption{LLM circle-type diagram for the scattering through the right boundary
in the unfolded picture. The vertical $D5$-brane corresponds to the
dot in the center of the circle. The line adjoining the center and
the circle corresponds to the boundary rep. The line segments to the
left from the boundary line correspond to the left reps, while the
line segments to the right from the boundary correspond to the right
reps. The phase is increasing towards the boundary for left and right
reps. Here a) is the scattering of the right rep through the boundary,
b) is the scattering of two left states in the bulk and c) is is the
scattering of the left rep through the boundary. The gray line segments
do not participate in the scattering.}
\label{fig_Kv_LLM}\par\end{centering}
\end{figure}

The achiral reflection in the unfolded picture may be understood as a scattering
through the achiral boundary and the choice of phases in (\ref{K_ach})
may be easily read from the LLM \cite{LLM} circle-type diagram (figure
\ref{fig_Kv_LLM}). The maps (\ref{K_ach_R}) and (\ref{K_ach_L}) in the unfolded
picture become
\begin{equation}
\Kappa^{\text{unf}}:1\otimes\check{\boxslash}\otimes\widetilde{\boxslash} \mapsto \boxslash\otimes\check{\boxslash}\otimes1,\label{K_ach_R_unf}
\end{equation}
and
\begin{equation}
\Kappa^{\text{unf}}:\boxslash\otimes\check{\boxslash}\otimes1 \mapsto 1\otimes\check{\boxslash}\otimes\widetilde{\boxslash},\label{K_ach_L_unf}
\end{equation}
respectively; thus folded and unfolded achiral reflection matrices are related to
each other as $\Kappa=\rm T\cdot\Kappa^{\text{unf}}$,
where T is the specialization of the folding map (\ref{T}) in
which the boundary carries an irreducible $\g_+$ rep (as opposed, more generally,
to a $\g_+$-reducible $\g_L\times \g_R$ rep). Note that the reflection coefficients
$k_3$, $k_4$ and $k_8$ acquire an extra minus sign in the unfolded picture (because of
graded permutation of two fermionic states).

Now we are ready to consider the explicit realization of the Yangian
(\ref{Y(gxg,g)}) for the vertical $D5$-brane. The boundary $\check{\boxslash}$
in this case is an evaluation irrep of the achiral twisted Yangian with rapidity zero.
We consider the folded picture first. All complementary central charges
(\ref{c_bar}) have zero eigenvalues on all (bulk and boundary) reps, hence
are trivial in this case and do not need to be considered.
Thus only the co-products of non-central charges in (\ref{Y(gxg,g)}) contribute.
Writing their action on the tensor product
$\boxslash\circ\widetilde{\boxslash}\otimes\check{\boxslash}$, we have, for example,
\begin{align}
\Delta\widetilde{\mathbb{R}}_{\check{a}}^{\enskip\check{b}} & =\check{\mathbb{R}}_{\check{a}}^{\enskip\check{c}}\circ1\otimes1-1
\circ\check{\mathbb{R}}_{\check{a}}^{\enskip\check{c}}\otimes1\nonumber \\
 & \qquad+\frac{1}{2}\mathbb{R}_{\check{a}}^{\enskip\check{c}}
 \circ\mathbb{R}_{\check{c}}^{\enskip\check{b}}\otimes1-\frac{1}{2}
 \mathbb{R}_{\check{c}}^{\enskip\check{b}}\circ
 \mathbb{R}_{\check{a}}^{\enskip\check{c}}\otimes1-\frac{1}{2}
 \mathbb{G}_{\check{a}}^{\enskip\check{\gamma}}\circ
 \mathbb{Q}_{\check{\gamma}}^{\enskip\check{b}}\otimes1-\frac{1}{2}
 \mathbb{Q}_{\check{\gamma}}^{\enskip\check{b}}\circ
 \mathbb{G}_{\check{a}}^{\enskip\check{\gamma}}\otimes1\nonumber \\
 & \qquad+\frac{1}{4}\delta_{\check{a}}^{\check{b}}\,
 \mathbb{G}_{\check{c}}^{\enskip\check{\gamma}}
 \circ\mathbb{Q}_{\check{\gamma}}^{\enskip\check{c}}\otimes1+
 \frac{1}{4}\delta_{\check{a}}^{\check{b}}\,
 \mathbb{Q}_{\check{\gamma}}^{\enskip\check{c}}\circ
 \mathbb{G}_{\check{c}}^{\enskip\check{\gamma}}\otimes1\nonumber \\
 & \qquad+\frac{1}{2}\mathbb{R}_{\check{a}}^{\enskip\check{c}}
 \circ1\otimes\mathbb{R}_{\check{c}}^{\enskip\check{b}}-\frac{1}{2}
 \mathbb{R}_{\check{c}}^{\enskip\check{b}}\circ1\otimes
 \mathbb{R}_{\check{a}}^{\enskip\check{c}}+\frac{1}{2}
 \mathbb{G}_{\check{a}}^{\enskip\check{\gamma}}\circ1\otimes
 \mathbb{Q}_{\check{\gamma}}^{\enskip\check{b}}+\frac{1}{4}
 \mathbb{Q}_{\check{\gamma}}^{\enskip\check{b}}\circ1\otimes
 \mathbb{G}_{\check{a}}^{\enskip\check{\gamma}}\nonumber \\
 & \qquad-\frac{1}{4}\delta_{\check{a}}^{\check{b}}\,
 \mathbb{G}_{\check{c}}^{\enskip\check{\gamma}}\circ1\otimes
 \mathbb{Q}_{\check{\gamma}}^{\enskip\check{c}}-\frac{1}{4}
 \delta_{\check{a}}^{\check{b}}\,
 \mathbb{Q}_{\check{\gamma}}^{\enskip\check{c}}\circ1\otimes
 \mathbb{G}_{\check{c}}^{\enskip\check{\gamma}}\nonumber \\
 & \qquad-\frac{1}{2}\,1\circ\mathbb{R}_{\check{a}}^{\enskip\check{c}}\otimes
 \mathbb{R}_{\check{c}}^{\enskip\check{b}}+\frac{1}{2}\,1\circ
 \mathbb{R}_{\check{c}}^{\enskip\check{b}}\otimes
 \mathbb{R}_{\check{a}}^{\enskip\check{c}}-\frac{1}{2}\,1\circ
 \mathbb{G}_{\check{a}}^{\enskip\check{\gamma}}\otimes
 \mathbb{Q}_{\check{\gamma}}^{\enskip\check{b}}-\frac{1}{4}\,1\circ
 \mathbb{Q}_{\check{\gamma}}^{\enskip\check{b}}\otimes
 \mathbb{G}_{\check{a}}^{\enskip\check{\gamma}}\nonumber \\
 & \qquad+\frac{1}{4}\delta_{\check{a}}^{\check{b}}\,1\circ
 \mathbb{G}_{\check{c}}^{\enskip\check{\gamma}}\otimes
 \mathbb{Q}_{\check{\gamma}}^{\enskip\check{c}}+\frac{1}{4}
 \delta_{\check{a}}^{\check{b}}\,1\circ
 \mathbb{Q}_{\check{\gamma}}^{\enskip\check{c}}\otimes
 \mathbb{G}_{\check{c}}^{\enskip\check{\gamma}},
\end{align}
and similarly for other charges. Thus the co-products in (\ref{Y(gxg,g)})
may be cast in the form
\begin{align}
\Delta\widetilde{\mathbb{J}}^{\check{A}} & =\Bigl(\hat{\mathbb{J}}^{\check{A}}\circ1-
1\circ\hat{\mathbb{J}}^{\check{A}}+\frac{1}{2}f_{\;\,\check{B}\check{C}}^{\check{A}}
(\mathbb{J}^{\check{B}}\circ\mathbb{J}^{\check{C}})\Bigr)\otimes1+\frac{1}{2}f_{\;\,\check{B}\check{C}}^{\check{A}}\Bigl(
 \mathbb{J}^{\check{B}}\circ1-1\circ\mathbb{J}^{\check{B}}\Bigr)\otimes\mathbb{J}^{\check{C}},\label{Y_Kv}
\end{align}
revealing the factorization (\ref{Kv_factor}) explicitly. Here the
first line corresponds to the Yangian of the $S$-matrix in (\ref{Kv_factor})
\begin{align}
\Delta\widetilde{\mathbb{J}}^{\check{A}}\Big|_{\mathcal{S}}
  & =\Bigl(\hat{\mathbb{J}}^{\check{A}}\circ1-1\circ\hat{\mathbb{J}}^{\check{A}}
       +\frac{1}{2}f_{\;\,\check{B}\check{C}}^{\check{A}}(\mathbb{J}^{\check{B}}\circ\mathbb{J}^{\check{C}})\Bigr)\otimes1,\label{Y_S}
\end{align}
and was explicitly spelled out in (\ref{Y_cop_D5h}), while the terms in
the second line originate from the achiral reflection matrices. Therefore the
part of the Yangian charges governing the achiral $\Kappa$-matrices is
\begin{align}
\Delta\widetilde{\mathbb{J}}^{\check{A}}{\Big|}_{\text {\large $\kappa$}}
& =\Bigl(\hat{\mathbb{J}}^{\check{A}}\circ1-1\circ\hat{\mathbb{J}}^{\check{A}}\Bigr)\otimes1
+\frac{1}{2}f_{\;\,\check{B}\check{C}}^{\check{A}}\Bigl(\mathbb{J}^{\check{B}}\circ1-
1\circ\mathbb{J}^{\check{B}}\Bigr)\otimes\mathbb{J}^{\check{C}}.\label{Y_K}
\end{align}

Once again we meet minus signs that need to be understood.
The origin of the minus sign in front of the level-1 charge in (\ref{Y_S})
was discussed in section 4.1, and its physical interpretation is almost
the same as for the horizontal reflection. The difference is that now both momenta and
rapidities of left and right reps are facing in the \textit{same} direction (towards
the boundary) in the initial configuration i.e. for the incoming state;
see left side of figure \ref{fig_Kv}) but the scattering always follows after
the reflection as seen from (\ref{Kv_factor}). Thus the $S$-matrix in (\ref{Kv_factor})
acts on the state with momentum and rapidity \textit{reversed} with respect to the
initial configuration.

In order to understand the origin of the minus sign in front of the level-1 charge
in (\ref{Y_K}) it is better to consider the unfolded
picture of the achiral reflection (the right side of figure \ref{fig_Kv}) first.
As for the horizontal case, the reflection in the unfolded picture may be
thought of as achiral scattering through the boundary state; this was nicely
shown in figure \ref{fig_Kv_LLM}. The Yangian charges (\ref{Y_K}) in the unfolded picture become
\begin{align}
\Delta\widetilde{\mathbb{J}}^{\check{A}}\Big|_{\text {\large $\kappa$}}^{\text{unf}}
  & =\hat{\mathbb{J}}^{\check{A}}\otimes1\otimes1+1\otimes1\otimes\hat{\mathbb{J}}^{\check{A}}
    +\frac{1}{2}f_{\;\,\check{B}\check{C}}^{\check{A}}\Bigl(\mathbb{J}^{\check{B}}\otimes
\mathbb{J}^{\check{C}}\otimes1+1\otimes\mathbb{J}^{\check{B}}\otimes\mathbb{J}^{\check{C}}\Bigr).
\end{align}
The minus sign in front of level-1 charge was absorbed by the unfolding map T$^{-1}$,
while the minus sign in front of the two-site term was absorbed into $f^{\check{A}}_{\;\,\check{B}\check{C}}$
using the antisymmetry under exchange of  $\check{B}$ and $\check{C}$. Thus the achiral
scattering through the boundary is governed by a Yangian symmetry equivalent to the bulk Yangian (\ref{Y(g)})
up to the different underlying tensor space structures. But here lies the most important
feature of the achiral scattering. In contrast to the horizontal case, the right rep in
the unfolded picture has momentum and rapidity pointing in \textit{opposite} directions,
as may be seen in the right side of figure \ref{fig_Kv}. Thus the minus sign in front of
the level-1 charge in (\ref{Y_K}) effectively reverses the rapidity of the right rep in
the folded picture, where it is pointing the same direction as the momentum.

We have checked that the Yangian charges (\ref{Y_Kv}) commute with the reflection matrix
$\mathcal{K}^v$ found in \cite{CRY}. Hence reflection from the vertical $D5$-brane may be
viewed in two ways: as a single reflection matrix $\mathcal{K}^v$ which is governed by
the Yangian charges (\ref{Y_Kv}), or as a factorized reflection (\ref{Kv_factor}) which is governed by
the Yangian charges (\ref{Y_S}) and (\ref{Y_K}). Either way the result is the same, as required.


\section{Discussion}

In this paper we have considered the achiral boundary conditions which result in
symmetry-breaking of the form $G\times G \rightarrow G$, in the contexts both of
the bosonic PCM and the AdS/CFT planar light-cone scattering theory. We have uncovered
the hidden boundary symmetry, which is an achiral twisted Yangian $Y(\g\times \g,\g)$
and differs subtly from a simple embedding of $Y(\g)$ in $Y(\g \times \g)$.
The difference is made precise by a mechanism which folds bulk, whole-line
scattering processes into boundary processes.

In the AdS/CFT case we have
considered reflection from both the `horizontal' and `vertical' $D5$-branes,
constructed the corresponding achiral twisted Yangian, and verified that this
is indeed the hidden symmetry of the known boundary $S$-matrices \cite{CRY}.
In the vertical case we have explained how these naturally factorize into a
composition of three events: the achiral reflection of the right rep of the
bulk magnon, bulk scattering of two left reps of the same magnon, and finally
achiral reflection of the left rep (mnemonically `right--bulk--left',
or equivalently, via the Yang-Baxter property, `left--bulk--right').
This achiral reflection is described by a new $\Kappa$-matrix, non-trivial
only for non-singlet boundaries and thus, in the AdS/CFT context, for the
`vertical' $D5$-brane only. For the bosonic PCM it might be interesting
to construct the full boundary theory, with its set of non-singlet boundary
bound states and their associated scattering theory.

Finally, it is striking that the two classes of integrable boundary condition of
the bosonic PCM, whose symmetries are generalized twisted Yangians based on
the two classes of compact symmetric spaces, should each find a realization
in AdS/CFT. It seems to us that this is further evidence for the view \cite{MacKayY}
that such algebras are the natural boundary symmetries that accompany bulk Yangian
symmetry. It will be interesting to discover whether they play a role in the eventual
full covariant quantization of AdS/CFT.

\textbf{Acknowledgements}. The authors would like to thank
Diego Correa, Alessandro Torrielli and Charles Young for useful
discussions and comments, and the UK EPSRC for funding under
grant EP/H000054/1.


\newcommand{\nlin}[2]{\href{http://xxx.lanl.gov/abs/nlin/#2}{\tt nlin.#1/#2}}
\newcommand{\hepth}[1]{\href{http://xxx.lanl.gov/abs/hep-th/#1}{\tt hep-th/#1}}
\newcommand{\arXivid}[1]{\href{http://arxiv.org/abs/#1}{\tt arXiv:#1}}
\newcommand{\Math}[2]{\href{http://xxx.lanl.gov/abs/math.#1/#2}{\tt math.#1/#2}}

\end{document}